\newif\iffigs
  \figsfalse
\documentclass[12pt]{article}
\usepackage{colordvi}
\usepackage{latexsym,amssymb}
\iffigs
    \usepackage{graphicx}
    \input{epsf}
\else
\fi
\newcommand{\ft}[2]{{\textstyle\frac{#1}{#2}}}
\newsavebox{\uuunit}
\sbox{\uuunit}
                             {\setlength{\unitlength}{0.825em}
                                      \begin{picture}(0.6,0.7)
                                                                  \thinlines
                                                                  \put(0,0){\line(1,0){0.5}}
                                                                  \put(0.15,0){\line(0,1){0.7}}
                                                                  \put(0.35,0){\line(0,1){0.8}}
                                                                 \multiput(0.3,0.8)(-0.04,-0.02){10}{\rule{0.5pt}{0.5pt}}
                                      \end {picture}}

\makeatletter \@addtoreset{equation}{section} \makeatother

\newcommand{\SO}{\mathop{\rm SO}}
\newcommand{\U}{\mathop{\rm {}U}}

\newcommand{\Sp}{\mathop{\rm {}Sp}}

\def\bfone{\relax{\rm 1\kern-.35em 1}}
\def\hat{\widehat}

\def\e{\epsilon}
\newcommand{\so}{\mathfrak{so}}
\newcommand{\su}{\mathfrak{su}}

\newcommand{\sym}{\mathfrak{sp}}
\newcommand{\uu}{\mathfrak{u}}

\usepackage{ifpdf}
\ifpdf
   \pdfoutput=1
   \pdftrue
  \usepackage[pdftex]{hyperref}
\else
   \usepackage{cite}
\fi
\begin{document}
\begin{titlepage}
\begin{flushright}
DISTA-2008 \\
\end{flushright}
\vskip 1.5cm
\begin{center}
{\LARGE \bf Superstrings on $\mathrm{AdS}_4 \times \mathbb{CP}^3$  from Supergravity
} \\
 \vfill
{\bf Riccardo D'Auria$^{~1}$, Pietro Fr{\'e}$^{~2}$,
\\ \vskip .2cm
Pietro Antonio Grassi$^{~3}$, and Mario Trigiante$^{~1}$}
\vfill {
$^1$ Dipartimento di Fisica Politecnico di Torino,\\ C.so Duca degli
Abruzzi, 24,
I-10129 Torino, Italy, \\
$^2$ Dipartimento di Fisica Teorica, Universit{\`a} di Torino, \\
$\&$ INFN -
Sezione di Torino\\
via P. Giuria 1, I-10125 Torino, Italy\\
\vskip .3cm
$^3$ 
{ DISTA, Universit\`a del Piemonte Orientale, }\\
{ Via Bellini 25/G,  Alessandria, 15100, Italy
$\&$ INFN - Sezione di
Torino
\vskip .3cm
%\\
%{CERN, Theory Unit, CH-1211 Geneve, 23, Switzerland}\\
}}
\end{center}
\vfill
\begin{abstract}
We derive from a general formulation of pure spinor string theory on type IIA
backgrounds the specific form of the action for the $AdS_4 \times {\mathbb CP}^3$ background.
We provide a complete geometrical characterization of the structure of the superfields
involved in the action.
\end{abstract}
\vfill
\vspace{1.5cm}
\vspace{2mm} \vfill \hrule width 3.cm {\footnotesize $^ \dagger $
This work is supported in part by the European Union RTN contract
MRTN-CT-2004-005104 and by the Italian Ministry of University (MIUR) under
contracts PRIN 2005-024045 and PRIN 2005-023102}
\end{titlepage}
%%%%%%%%%%%%%%%%%%%%%%%%%%%%%%%%%%%%%%%%%%
\tableofcontents
%%%%%%%%%%%%%%%%%%%%
\section{Introduction}

\def\N{\mathcal{N}}

The recent developments on the duality between $\N=6$ superconformal Chern-Simons
theory in three dimensions and superstrings moving on ${\rm AdS}_4 \times {\mathbb P}^3$
\cite{{Aharony:2008ug}, Benna:2008zy,Schwarz:2004yj,Bagger:2007vi,Bagger:2006sk,Gustavsson:2007vu,Gustavsson:2008dy,Distler:2008mk,Lambert:2008et}
have prompted the study of superstrings on ${\rm Osp}({\mathcal N}|4)$ backgrounds
\cite{Arutyunov:2008if, Stefanski:2008ik,Fre:2008qc, 
Bonelli:2008us}. The main issue is of course the integrability
of the system and  this has been already studied in a series of papers \cite{Minahan:2008hf,Gaiotto:2008cg,Bak:2008cp,Gromov:2008bz,Gromov:2008qe,Ahn:2008aa,Ahn:2008hj,Lee:2008ui,Astolfi:2008ji,Chen:2008qq,Ahn:2008gd,Grignani:2008te}. On the other side, one would like also to
consider the string theory in a framework where all symmetries are manifest and which
takes the RR fields of the background  properly into account. In \cite{Bonelli:2008us}, the 
limit for large RR fields is analyzed and it has been shown the relation with a topological model 
on the Grassmannian ${\rm Osp}({\mathcal 6}|4) / {\mathrm SO}(6) \times {\mathrm Sp}(4)$. The 
exactness of the background is also discussed in \cite{Bonelli:2008us}.  

The pure spinor formalism is well suited to the present situation
and in a previous paper  \cite{Fre:2008qc} two of the present
authors provided the pure spinor version of the ${\rm AdS}_4 \times
{\mathbb P}^3$ sigma model, described as the coset space ${\rm
Osp}(6|4)/{\rm SO}(1,3) \times {\rm U}(3)$. Furthermore, the four
authors published another paper \cite{D'Auria:2008ny} where a
systematic study of pure spinor superstring on type IIA backgrounds
has been completely performed.  This analysis has been based on the
previous studies by Berkovits and Howe \cite{Berkovits:2001ue}, by
Oda and Tonin \cite{Oda:2001zm} and on the geometric (a.k.a.
rheonomic) formulation of supergravity \cite{Castellani:1991et}.
There it has been shown how to derive from the geometrical
formulation of supergravity (in type IIA case) the pure spinor sigma
model and the relative pure spinor constraints \cite{Psconstra}. It
has been proved that the action is BRST invariant and, only in the
case of type IIA, has a peculiar structure since it can be written
in terms of four pieces which are the Green-Schwarz action, a
$Q$-exact piece, a $\bar Q$-exact piece and a $Q\bar Q$-exact piece.
This allows us to derive the complete expression of the sigma model
where all superfields are made explicit. One of the advantages of
the geometrical formulation of supergravity is that it provides a
superspace framework where all bosonic fields are extended to be
superfields and the rheonomic conditions ensure the integrability of
the extension, leading to the correct field content. The advantage
stays in the fact that one can very easily read off the sigma model
action in terms of the background solution.
 As an example, here we derive of the pure spinor sigma model
for the ${\rm AdS}_4 \times {\mathbb P}^3$ background.

In this case we have to take into account the RR field stregths
${\bf G}^{[2]}$ and ${\bf G}^{[4]}$ which are respectively
proportional to the K\"ahler form on ${\mathbb P}^3$ and to the
Levi-Civita invariant tensor in ${\rm AdS}_4$. This background has
24 Killing spinors parametrized by the combinations $\chi_x \otimes
\eta^A$ where $\chi_x$ are the Killing spinors of $\mathrm{AdS}_4$
and $\eta^A$ are the 6 Killing spinors of ${\mathbb P}^3$.
Therefore, it is convenient to use a superspace with 24 fermionic
coordinates. Now, the problem is whether this superspace is
sufficient to provide a complete description of the supergravity
states and, whether the vertex operators constructed in terms of
this superspace describe on-shell ${\rm AdS}_4 \times {\mathbb
P}^3$-supergravity fluctuations.  It is established that all
supergravity models with more than 16 supercharges are described by
an on-shell superspace, since an auxiliary-field formulation does
not exist, and therefore we expect that the 24-extended superspace
is sufficient for the present formulation. There is also another
aspect to be noticed: the formulation of GS superstrings on the same
coset has been studied extensively in \cite{Arutyunov:2008if} and it
has been argued that 24 fermions are indeed sufficient to formulate
the model. Indeed, $\kappa$-symmetry removes exactly 8 fermions
leading to a supersymmetric model. In our case, $\kappa$-symmetry is
replaced by BRST symmetry plus pure spinor constraints, so that we
have to check whether the pure spinors satisfying the new
constraints \cite{Psconstra} cancel the central charge. In fact, we
will see that by reducing the spinor space from 32 dimensions to the
24 dimensions adapted to the present background, there exists a
solution of the pure spinor constraints with only 14 degrees of
freedom, matching the bosonic and fermionic degrees of freedom.

In addition, by means of the formalism constructed in
\cite{D'Auria:2008ny}, we provide and explicit expression for the
sigma model where all couplings are exhibited. We devote a
particular attention on the  quartic part of the action for the
ghosts.\par The paper is organized as follows. In Section \ref{s2}
we review the description of Type IIA supergravity in terms of its
Free Differential Algebra (FDA) in the string frame and the
corresponding rheonomic parametrization. In section \ref{s3} we
describe the compactification of type IIA on $\mathrm{AdS}_4 \,
\times \, {\mathbb{P}}^3$. Finally in section \ref{s4} we give the
complete pure spinor  superstring action on $\mathrm{AdS}_4 \,
\times \, {\mathbb{P}}^3$. The reader is referred to the appendices
for a definition of the $D=4$ and $D=6$ spinor conventions and for
some useful formulae.

%%%%%%%%%%%%%%%%%%%%%%%%%%%%%%%%%%%%%%%%%%%
\section{Summary of Type IIA Supergravity and of its FDA}
\label{s2} In order to pursue our programme we have to consider the
structure of the Free Differential Algebra of type IIA supergravity,
the rheonomic parametrization of its curvatures and the
corresponding field equations that are the integrability conditions
of such rheonomic parametrizations. All these necessary ingredients
were recently determined in \cite{D'Auria:2008ny}. In this section,
we summarize those results collecting all the
 items for our subsequent discussion.
\subsection{Definition of the curvatures}
The p-forms entering the FDA of the type IIA theory are listed below:
\par
\vskip 0.2cm
\begin{centering}
\begin{tabular}{|c|c|c|c|c|c|}
  \hline
  Form & degree p & f(ermion)/b(oson) & Name & String Sector & Curvature \\
  \hline
  $\omega^{\underline{ab}}$ & 1 & b & spin connection & NS-NS & $R^{\underline{ab}}$ \\
  $V^{\underline{a}}$ & 1 & b & Vielbein & NS-NS & $T^{\underline{a}}$ \\
  $\psi_{L/R}$ & 1 & f & gravitino & NS-R & $\rho_{L/R}$ \\
  $\mathbf{C}^{[1]}$ & 1 & b & RR 1-form & R-R & $\mathbf{G}^{[2]}$ \\
  $\varphi$ & 0 & b & dilaton & NS-NS & $\mathbf{f}^{[1]}$ \\
  $\chi_{L/R}$ & 0 & f & dilatino & NS-R & $\nabla\chi_{L/R}$ \\
  $\mathbf{B}^{[2]}$ & 2 & b & Kalb-Ramond field & NS-NS & $\mathbf{H}^{[3]}$ \\
  $\mathbf{C}^{[3]}$ & 3 & b & RR 3-form & R-R & $\mathbf{G}^{[4]}$ \\
\hline
\end{tabular}
\end{centering}
\par
\vskip 0.2cm
The explicit definition of the FDA curvatures, constructed with the
above fields is displayed below:
\begin{eqnarray}
R^{\underline{ab}} & \equiv & d\omega^{\underline{ab}} \, - \, \omega^{\underline{ac}}\, \wedge \,
\omega^{\underline{cb}}\label{2acurva}\\
T^{\underline{a}} & \equiv & \mathcal{D} \, V^{\underline{a}} \, - \,
{\rm i} \, \ft 12 \left(\overline{\psi}_L \, \wedge \,\Gamma^{\underline{a}} \, \psi_L \, + \,
\overline{\psi}_R \, \wedge \,\Gamma^{\underline{a}} \, \psi_R \right)
\label{2atorsio}\\
\rho_{L,R} & \equiv & \mathcal{D}\psi_{L,R} \, \equiv \, d\psi_{L,R} \, - \,
\ft 14 \omega^{\underline{ab}} \, \wedge \, \Gamma_{\underline{ab}} \,\psi_{L,R} \label{a2curlgrav}\\
\mathbf{G}^{[2]} & \equiv & d\mathbf{C}^{[1]} \, + \, \exp\left[ - \, \varphi \right]
\, \overline{\psi}_R \, \wedge \, \psi_L \label{F2defi}\\
\mathbf{f}^{[1]} & \equiv & d\varphi \label{dilacurva}\\
\nabla \chi_{L/R} &\equiv& \, d\chi_{L,R} \, - \,
\ft 14 \omega^{\underline{ab}} \, \wedge \, \Gamma_{\underline{ab}} \,\chi_{L,R} \label{a2curldil}
\end{eqnarray}
\begin{eqnarray}
\mathbf{H}^{[3]} & = & d\mathbf{B}^{[2]} \, + \, {\rm i} \,
\left(\overline{\psi}_L \, \wedge \,
\Gamma_{\underline{a}} \, \psi_L \,
- \, \overline{\psi}_R \, \wedge \,\Gamma_{\underline{a}} \,
\psi_R \right) \, \wedge \, V^{\underline{a}}
\label{KRcurva}\\
\mathbf{G}^{[4]} & =
& d\mathbf{C}^{[3]} \, + \, \mathbf{B}^{[2]} \, \wedge \,
d\mathbf{C}^{[1]}\, \nonumber\\
&& - \, \ft 12 \, \exp\left [- \, \varphi \right]\,\left(\overline{\psi}_L \, \wedge \,
\Gamma_{\underline{ab}} \, \psi_R \,
+ \, \overline{\psi}_R \, \wedge \,\Gamma_{\underline{ab}} \, \psi_L
\right)\, \wedge \, V^{\underline{a}} \, \wedge \, V^{\underline{b}} \label{Gcurva}
\end{eqnarray}
The $0$--form dilaton $\varphi$ appearing in eq. (\ref{F2defi})
introduces a dynamic coupling constant. Furthermore, as mentioned in
the table, $V^{\underline{a}}$, and $\omega^{\underline{ab}}$
respectively denote the vielbein and the spin connection, which
together with the gravitino  $\psi_{L/R}$ complete the multiplet of
$1$-forms gauging the type IIA super Poincar\'e algebra in $D=10$.
The two fermionic $1$-forms $\psi_{L/R}$ are Majorana-Weyl spinors
of opposite chirality:
\begin{equation}
  \Gamma_{11} \, \psi_{L/R} \, = \, \pm \, \psi_{L/R}\,.
\label{psiLR}
\end{equation}
The flat metric $\eta_{\underline{ab}}\, = \, \mbox{diag}(+,-,\dots,-)$
is the mostly minus one and $\Gamma_{11}$ is hermitian and
squares to the the identity $\Gamma_{11}^2=\mathbf{1}$.
\subsection{Rheonomic parametrizations of the curvatures in
the string frame} \label{stringframerheo} As explained in
\cite{D'Auria:2008ny} the form of the rheonomic parametrization
required in order to construct the pure spinor action of
superstrings is that corresponding to the string frame and not that
corresponding to the Einstein frame. This parametrization was
derived in \cite{D'Auria:2008ny} and it is formulated in terms of a
certain set of tensors, which involve both the supercovariant field
strengths
$\mathcal{G}_{\underline{ab}},\mathcal{G}_{\underline{abcd}}$ of the
Ramond-Ramond $p$-forms and also bilinear currents in the dilatino
field $\chi_{L/R}$. The needed tensors are those listed below:
\begin{eqnarray}
{\mathcal{M}}_{\underline{ab}} & = & \Big( \ft 18 \, \exp[
\varphi] \, \mathcal{G}_{\underline{ab}} \, + \, \ft {9}{64} \,
\overline{\chi}_R \, \Gamma_{\underline{ab}} \, \chi_L \Big)
\nonumber\\
{\mathcal{M}}_{\underline{abcd}} & = &  - \, \ft 1{16} \, \exp[
\varphi] \, \mathcal{G}_{\underline{abcd}}
%\,\nonumber\\
 %&&
 - \, \ft{3}{256}  \,  \overline{\chi}_L \,
\Gamma_{\underline{abcd}} \, \chi_R \nonumber\\
\mathcal{N}_0 \, & = &  \ft 34 \, \overline{\chi}_L\, \chi_R \nonumber\\
\mathcal{N}_{\underline{ab}}&=&
\ft 14  \, \exp[ \varphi] \, \mathcal{G}_{\underline{ab}} \, + \, \ft{9}{32} \,
\overline{\chi}_R \, \Gamma_{\underline{ab}}
\, \chi_L=2\,{\mathcal{M}}_{\underline{ab}} \nonumber\\
%\mathcal{N}_{\underline{abc}}^{\pm} \, & = & \ft{1}{64} \,
%\overline{\chi}_{L/R} \, \Gamma_{\underline{abc}}
%\, \chi_{L/R} \nonumber\\
\mathcal{N}_{\underline{abcd}} \, &= &  \ft 1{24} \, \exp[ \varphi] \,
\mathcal{G}_{\underline{abcd}}
%\,\nonumber\\
%&&
+ \, \ft{1}{128} \, \overline{\chi}_R \,\Gamma_{\underline{abcd}} \,
\chi_L = - \ft 23 {\mathcal{M}}_{\underline{abcd}}\,.
\label{Mntensors}
\end{eqnarray}
The above tensors are conveniently assembled into the following
spinor matrices
\begin{eqnarray}
\mathcal{M}_\pm &=&{\rm i} \, \left(\mp
\mathcal{M}_{\underline{ab}} \,
  \Gamma^{\underline{ab}} \, + \,  \mathcal{M}_{\underline{abcd}} \,
  \Gamma^{\underline{abcd}}\right)
\label{Mmatrapm}\\
\mathcal{N}^{(even)}_{\pm} & = & \mp \,\mathcal{N}_0 \, \mathbf{1}
\, + \, \mathcal{N}_{\underline{ab}} \, \Gamma^{\underline{ab}} \,
\mp \, \mathcal{N}_{\underline{abcd}} \, \Gamma^{\underline{abcd}}
\label{pongo}\\
 \mathcal{N}^{(odd)}_{\pm} & = &\pm \ft
i3\,f_{\underline{a}}\,\Gamma^{\underline{a}}\pm\ft{1}{64}\,\overline{\chi}_{R/L}
\, \Gamma_{\underline{abc}}\,
\chi_{R/L}\,\Gamma^{\underline{abc}}-\ft i{12}\,\mathcal{H}_{\underline{abc}}\,\Gamma^{\underline{abc}}\label{pongo2}\\
\mathcal{L}^{(odd)}_{a\,\pm}&=&
\mathcal{M}_{\mp}\,\Gamma_{\underline{a}}\,\,;\,\,\,\mathcal{L}^{(even)}_{a\,\pm}=\mp\ft
38\,\mathcal{H}_{\underline{abc}}\,\Gamma^{\underline{bc}}\,.
\end{eqnarray}
\par In terms of these objects the rheonomic parametrizations of
the curvatures, solving the Bianchi identities can be written as
follows:
\paragraph{Bosonic curvatures}
\begin{eqnarray}
T^{\underline{a}} & = & 0 \label{nullatorsioSF}\\
R^{\underline{ab}} & = & R^{\underline{ab}}{}_{\underline{mn}} \,
V^{\underline{m}} \,\wedge \, V^{\underline{n}}\, + \,
\overline{\psi}_R\,{\Theta}^{\underline{ab}}_{\underline{m}|L}\,
\wedge \, V^{\underline{m}}\, + \,\overline{\psi}_L\,
{\Theta}^{\underline{ab}}_{\underline{m}|R} \, \wedge \, V^{\underline{m}}\, \nonumber\\
&& + \,{\rm i} \, \ft 34 \, \left( \overline{\psi}_L \, \wedge \,
\Gamma_{\underline{c}} \, \psi_L \, - \, \overline{\psi}_R \, \wedge
\, \Gamma_{\underline{c}} \, \psi_R \right)
\, \mathcal{H}^{\underline{abc}}\nonumber\\
&& \, + 2i\, \overline{\psi}_L \, \wedge \, \Gamma^{[\underline{a}}
\, \mathcal{M}_+ \, \Gamma^{\underline{b}]} \, \psi_R
\label{rheoRiemannSF}\\
\mathbf{H}^{[3]} & = & \mathcal{H}_{\underline{abc}} V^{\underline{a}} \, \wedge \, V^{\underline{b}} \, \wedge \,
V^{\underline{c}} \label{rheoHSF}\\
\mathbf{G}^{[2]} & = & \mathcal{G}_{\underline{ab}} V^{\underline{a}} \, \wedge \, V^{\underline{b}} \,
\, + \, {\rm i} \, \ft 32 \exp\left[  - \, \varphi \right] \, \left(\overline{\chi}_L  \,
\Gamma_{\underline{a}} \, \psi_L \,
+\, \overline{\chi}_R \,\Gamma_{\underline{a}} \, \psi_R
\right)\, \wedge \, V^{\underline{a}} \label{rheoFSF}\\
\mathbf{f}^{[1]} & = & f_{\underline{a}} V^{\underline{a}}  \, + \, \ft 32 \, \left(\overline{\chi}_R  \,
 \psi_L \, - \, \overline{\chi}_L \, \psi_R \right)\label{rheodilatonFSF}\\
 \mathbf{G}^{[4]} & = & \mathcal{G}_{\underline{abcd}} V^{\underline{a}} \, \wedge \, V^{\underline{b}} \, \wedge \,
V^{\underline{c}} \, \wedge \,
V^{\underline{d}}\label{rheoGSF} \nonumber\\
&&\, - \, {\rm i} \, \ft 12 \,
\exp[-\varphi] \, \left(\overline{\chi}_L \, \Gamma_{\underline{abc}} \, \psi_L \, - \,
\overline{\chi}_R \, \Gamma_{\underline{abc}} \, \psi_R \right) \,
\wedge \, V^{\underline{a}} \, \wedge \, V^{\underline{b}} \, \wedge \, V^{\underline{c}}
\end{eqnarray}
\paragraph{Fermionic curvatures}
\begin{eqnarray}
\rho_{L/R} & = &\rho^{L/R}_{\underline{ab}} \, V^{\underline{a}} \,
\wedge \, V^{\underline{b}} \,
+\mathcal{L}^{(even)}_{\underline{a}\,\pm}\,\psi_{L/R}\wedge
V^{\underline{a}}+\mathcal{L}^{(odd)}_{\underline{a}\,\mp}\,\psi_{R/L}\wedge
V^{\underline{a}} \, + \, \rho_{L/R}^{(0,2)}
 \label{rhoparaSF}\\
\nabla\, \chi_{L/R} & = & \mathcal{D}_{\underline{a}} \, \chi_{L/R}
\, V^{\underline{a}}
+\mathcal{N}^{(even)}_{\pm}\,\psi_{L/R}+\mathcal{N}^{(odd)}_{\mp}\,\psi_{R/L}\,.
\label{dechiparaSF}
\end{eqnarray}
Note that the components of the generalized curvatures along the
bosonic vielbeins do not coincide with their spacetime components,
but rather with their supercovariant extension. Indeed expanding
for example the four-form along the spacetime differentials one
finds that
\begin{eqnarray} \widetilde G_{\mu\nu\rho\sigma}
&\equiv&\mathcal{G}_{\underline{abcd}} V^{\underline{a}}_{\mu} \,
\wedge \, V^{\underline{b}}_{\nu} \, \wedge \,
V^{\underline{c}}_{\rho} \, \wedge \, V^{\underline{d}}_{\sigma} =
\partial_{[\mu}C_{\nu\rho\sigma]}^{[4]} + B_{[\mu\nu}^{[2]}\,\partial_\rho
C^{[1]}_{\sigma]}-\nonumber\\
&&-\frac{1}{2}\,e^{-\varphi}\,\left(\overline{\psi}_{L[\mu}\,\Gamma_{\nu\rho}\,\psi_{R\sigma]}+\overline{\psi}_{R[\mu}\,\Gamma_{\nu\rho}\,\psi_{L\sigma]}\right)
\nonumber\\
&&+ \, {\rm i} \, \ft 12 \, \exp[-\varphi] \,
\left(\overline{\chi}_L \, \Gamma_{[\mu\nu\rho} \, \psi_{L\sigma]}
\, - \, \overline{\chi}_R \, \Gamma_{[\mu\nu\rho} \,
\psi_{R\sigma]} \right)\nonumber\end{eqnarray} where $\widetilde
G$ is the supercovariant field strength.
\par In the
parametrization (\ref{rheoRiemannSF}) of the Riemann tensor we
have used the following definition:
\begin{eqnarray}
\Theta_{\underline{ab|c}L/R} &=& -i \Big( \Gamma_{\underline{a}}
\rho_{\underline{bc} R/L} + \Gamma_{\underline{b}}
\rho_{\underline{ca} R/L} - \Gamma_{\underline{c}}
\rho_{\underline{ab} R/L} \Big)\,.
\end{eqnarray}
Finally by $\rho^{(0,2)}_{L/R}$ we have denoted the fermion-fermion part of
the gravitino curvature whose explicit expression can be written in two
different forms, equivalent by Fierz rearrangement:
\begin{eqnarray}
\rho_{L/R}^{(0,2)}&=& \, \pm \, \ft{21}{32} \, \Gamma_{\underline{a}} \, \chi_{R/L} \, {\bar \psi}_{L/R} \, \wedge \,
\Gamma^{\underline{a}} \, \psi_{L/R} \nonumber\\
 && \mp \, \ft{1}{2560} \, \Gamma_{\underline{a_1a_2a_3a_4a_5}} \, \chi_{R/L} \,  \left (
 \overline{\psi}_{L/R} \,
 \Gamma^{\underline{a_1a_2a_3a_4a_5}} \, \psi_{L/R} \right )
 \label{rhoparaSF2}\\
 && \mbox{or} \nonumber\\
\rho_{L/R}^{(0,2)}&=& \, \pm \, \ft{3}{8} \, {\rm i}\, \psi_{L/R} \, \wedge \, {\bar \chi}_{R/L} \, \, \psi_{L/R}
  \, \pm \, \ft{3}{16} \, {\rm i}\, \Gamma_{\underline{ab}} \,\psi_{L/R} \, \wedge \, {\bar \chi}_{R/L} \, \,
  \Gamma^{\underline{ab}} \, \psi_{L/R}\,.
 \label{rhoparaSF3}
\end{eqnarray}
\subsection{Field equations of type IIA supergravity in the string frame}
\label{equefilde} The rheonomic parametrizations of the
supercurvatures displayed above imply, via Bianchi identities,
 a certain number of
constraints on the inner components of the same curvatures which
can be recognized as the field equations of type IIA supergravity
in the string frame. These are the equations that have to be solved
in constructing any specific supergravity background and read as
follows.
\par
We have an Einstein equation of
the following form:
\begin{eqnarray}
\mbox{\emph{$\mathcal{R}$}}_{\underline{ab}} & = & \hat T_{\underline{ab}}\left( f\right) \,  +
\, \hat T_{\underline{ab}}\left( \mathcal{G}_2\right) \,+ \, \hat T_{\underline{ab}}\left( \mathcal{H}\right)\, + \, \hat T_{\underline{ab}}
\left( \mathcal{G}_4\right) \label{Einsteinus}
\end{eqnarray}
where the stress-energy tensor on the right hand side are defined as
\begin{eqnarray}
\hat T_{\underline{ab}}\left( f\right) & = &\, - \, \mathcal{D}_{\underline{a}} \, \mathcal{D}_{\underline{b}}\varphi \,
+ \, \ft 89 \, \mathcal{D}_{\underline{a}} \, \varphi \, \mathcal{D}_{\underline{b}} \, \varphi
\, - \, \eta_{\underline{ab}} \left( \ft 16 \Box \, \varphi \,
+ \, \ft 59 \, \mathcal{D}^{\underline{m}} \, \varphi \, \mathcal{D}_{\underline{m}} \, \varphi
\right)\label{dialtostress}\\
\hat T_{\underline{ab}}\left( \mathcal{G}_2\right) & = &   \exp\left[2 \, \varphi
\right] \, \mathcal{G}_{\underline{ax}} \, \mathcal{G}_{\underline{by}} \,
\eta^{\underline{ab}} \label{Fstresso}\\
\hat T_{\underline{ab}}\left( \mathcal{H}\right)\, & = & \, - \,\exp\left[\ft 13 \, \varphi
\right] \, \left( \ft 98 \, \mathcal{H}_{\underline{axy}} \, \mathcal{H}_{\underline{bwt}} \,
\eta^{\underline{xw}} \, \eta^{\underline{yt}}
\, - \, \ft 18 \, \eta_{\underline{ab}} \, \mathcal{H}_{\underline{xyz}} \,
\mathcal{H}^{\underline{xyz}}\right) \label{Hstresso}\\
\hat T_{\underline{ab}}\left( \mathcal{G}_4\right) & = & \exp\left[2 \, \varphi
\right] \,\left( 6 \,\mathcal{G}_{\underline{ax_1x_2x_3}} \,
\mathcal{G}_{\underline{by_1y_2y_3}} \,
 \eta^{\underline{x_1y_1}} \,
\eta^{\underline{x_2y_2}}\, \eta^{\underline{x_3y_3}}\, - \, \ft 12
\, \eta_{\underline{ab}} \, \mathcal{G}_{\underline{x_1\dots x_4}}
\, \mathcal{G}^{\underline{x_1\dots x_4}}\right)\,.
\end{eqnarray}
Next we have the equations for the dilaton and the Ramond $1$-form:
\begin{eqnarray}
0 & = &\Box \, \varphi \, - \, 2 \,  f_{\underline{a}} \, f^{\underline{a}} \, + \, \ft 32 \,
\exp \left[2 \, \varphi \right]
\, \mathcal{G}^{\underline{x_1x_2}} \, \mathcal{G}_{\underline{x_1x_2}}\nonumber\\
& & + \, \ft 32 \,
\exp \left[2 \, \varphi \right]
\,
 \mathcal{G}^{\underline{x_1x_2x_3x_4}} \, \mathcal{G}_{\underline{x_1x_2x_3x_4}} \,
  + \, \ft 34 \,
\exp \left[\ft 43\, \varphi \right]
\,\mathcal{H}^{\underline{x_1x_2x_3}} \, \mathcal{H}_{\underline{x_1x_2x_3}}
\label{Rreq01}\\
0 & = & \mathcal{D}_{\underline{m}} \,\mathcal{G}^{\underline{ma}}\,
 - \, \ft 53 \,f^{\underline{m}} \, \mathcal{G}_{\underline{ma}}
 \, + \, 3 \, \mathcal{G}^{\underline{ax_1x_2 x_3}} \, \mathcal{H}_{\underline{x_1 x_2 x_3}}
 \label{G2equation}
\end{eqnarray}
and the equations for the NS $2$-form and for the RR $3$-form:
\begin{eqnarray}
0 & = & \mathcal{D}_{\underline{m}} \,\mathcal{H}^{\underline{mab}}
\, - \, \ft 23 \, f^{\underline{m}} \,
\mathcal{H}_{\underline{mab}}\nonumber\\
&&
 \, - \,  \exp\left[  \ft 43 \, \varphi \right]\, \left( 4 \,
\, \mathcal{G}^{\underline{x_1x_2 ab}} \, \mathcal{G}_{\underline{x_1 x_2}} \, - \, \ft {1}{24} \,
\epsilon^{\underline{abx_1 \dots x_8}} \,
\mathcal{G}_{\underline{x_1x_2x_3x_4}} \,\mathcal{G}_{\underline{x_5x_6x_7x_8}}\right)
\label{H3equa}\\
0 & = &
\mathcal{D}_{\underline{m}}\,\mathcal{G}^{\underline{ma_1 a_2a_3}} \,
+ \, \ft 13 \, f_{m} \, \mathcal{G}^{\underline{ma_1
a_2a_3}}\nonumber\\
&& + \, \exp\left[  \ft 23 \, \varphi \right]\, \left( \ft 32 \,
\mathcal{G}^{\underline{m}[\underline{a_1}} \,
H^{\underline{a_2a_3}]\underline{n}} \, \eta_{\underline{mn}} \,
  \, + \,\ft {1}{48} \, \epsilon^{\underline{a_1a_2a_3x_1 \dots x_7}} \mathcal{G}_{\underline{x_1x_2x_3x_4}}
 \, H_{\underline{x_5x_6x_7}}\right)\,.
\label{23formeque}
\end{eqnarray}
Any solution of these bosonic set of equations can be uniquely
extended to a full superspace solution involving $32$ theta variables
by means of the rheonomic conditions. The implementation of such a
fermionic integration is the \textit{supergauge completion}.

\section{Compactifications of type IIA on $\mathrm{AdS}_4 \, \times \,
{\mathbb{P}}^3$ }\label{s3} In this section we construct a
compactification of type IIA supergravity on the following direct
product manifold:
\begin{equation}
  \mathcal{M}_{10} \, = \, \mathrm{AdS}_4 \, \times \,
{\mathbb{P}}^3
\label{productspace}
\end{equation}
%\section{Gauge completion in  mini superspace}
The local symmetries of the effective theory on this background is
encoded in the supergroup $\mathrm{OSp(6|4)}$. The supergauge
completion of the $\mathrm{AdS}_4 \, \times \, {\mathbb{P}}^3$ space
consists in expressing the ten--dimensional superfields, satisfying
the rheonomic parametrizations in terms of the coordinates of the
\emph{mini-superspace} associated with this background, namely of
the 10 space-time coordinates $x^{\underline{\mu}}$ and the 24
fermionic ones $\theta$, parametrizing the preserved supersymmetries
only. This procedure relies on the representation of the
\emph{mini-superspace} in terms of the following super--coset
manifold
\begin{eqnarray}
\mathcal{M}^{10|24}&=&\frac{\mathrm{OSp(6|4)}}{\SO(1,3)\times\U(3)}\,.\label{supermanifoldamente}
\end{eqnarray}
The bosonic subgroup of $\mathrm{OSp(6|4)}$ is ${\rm
Sp}(4,\mathbb{R})\times \SO(6)$. The Maurer-Cartan 1--forms of
$\mathfrak{sp}(4,\mathbb{R})$ are denoted by $\Delta^{xy}$
($x,y=1,\dots, 4$), the $\mathfrak{so}(6)$ 1--forms are denoted by
$\mathcal{A}_{AB}$ ($A,B=1,\dots, 6$) while the (real) fermionic
1-forms are denoted by $\Phi^x_A$ and transform in the fundamental
representation of $\Sp(4,\mathbb{R})$ and in the fundamental
representation of $\SO(6)$. These forms satisfy the
$\mathrm{OSp(6|4)}$ Maurer-Cartan equations:
\begin{eqnarray}
d \Delta^{xy} + \Delta^{xz} \, \wedge \,\Delta^{ty} \, \epsilon_{zt}
&=&
 -\, 4 \, {\rm i}\,  e \,  {\Phi}_A^x \, \wedge \, {\Phi}_A^y, \nonumber \\
d {\mathcal{A}}_{AB} - e  {\mathcal{A}}_{AC}\, \wedge \,
{\mathcal{A}}_{CB} &=&
4 \, {\rm i}  {\Phi}_A^x \, \wedge \, {\Phi}_B^y \, \epsilon_{xy}\nonumber\\
d \Phi^x_A \, +  \, \Delta^{xy} \, \wedge \, \epsilon_{yz} \,
\Phi^z_A \, - \,
 e \, {\mathcal{A}}_{AB} \, \wedge \,\Phi^x_B &=&  0
\label{orfan26}
\end{eqnarray}
where
\begin{equation}
  \epsilon_{xy}= - \epsilon_{yx} \, = \, \left(\matrix{ 0 & 0 & 0 & 1 \cr 0 & 0 & -1 & 0 \cr 0 & 1 & 0 & 0 \cr
    -1 & 0 & 0 & 0 \cr  } \right)
\label{epsilon}
\end{equation}
The Maurer-Cartan equations are solved in terms of the super-coset
representative of (\ref{supermanifoldamente}). We rely for this
analysis on the general discussion in \cite{Fre:2008qc}. It is
convenient to express this solution in terms of the 1-forms
describing the on the bosonic submanifolds $\mathrm{AdS}_4\equiv
\frac{\Sp(4,\mathbb{R})}{\SO(1,3)}$, $\mathbb{P}^3\equiv
\frac{\SO(6)}{\U(3)}$ of (\ref{supermanifoldamente}) and 1--forms on
the fermionic subspace of (\ref{supermanifoldamente}). Let us denote
by  $B^{ab},\,B^{a}$ and by
$\mathcal{B}^{\alpha\beta},\,\mathcal{B}^{\alpha}$
 the connections and vielbein on the two bosonic subspaces
 respectively.
  The supergauge completion is
finally accomplished by expressing the $p$-forms satisfying the
rheonomic parametrization of the FDA in the mini-superspace. This
amounts to expressing them in terms of the 1--forms on
(\ref{supermanifoldamente}). The final expression of the $D=10$
fields will involve not only the bosonic 1--forms
$B^{ab},\,B^{a},\,\mathcal{B}^{\alpha\beta},\,\mathcal{B}^{\alpha}$,
but also the Killing spinors on the background. The latter play
indeed a spacial role in this analysis since they can be identified
with the fundamental harmonics of the cosets
$\mathrm{SO(2,3)/SO(1,3)}$ and $\mathrm{SO(6)/U(3)}$, respectively,
\cite{Fre':2006es}. Before writing the explicit solution we need to
discuss the Killing spinors on the $\mathrm{AdS}_4\times
\mathbb{P}^3$ background. \subsection{Killing spinors of the
$\mathrm{AdS_4}$ manifold} As anticipated, on of the  main items for
the construction of the supergauge completion is given by the
Killing spinors of anti de Sitter space. They can be constructed in
terms of the coset representative $\mathrm{L_B}$, namely in terms of
the fundamental harmonic of the coset $\mathrm{SO(2,3)/SO(1,3)}$.
\par
The defining equation is given by:
\begin{equation}
\nabla^{\mathrm{Sp(4)}}\, \chi_x \, \equiv \, \left( d  \, - \, \ft
14 \, B^{ab} \, \gamma_{ab} \, - \,2\, e \, \gamma_a \, \gamma_5 \,
 B^a \,\right) \, \chi_x \, = \, 0
\label{d4Killing}
\end{equation}
and states that the Killing spinor is a covariantly constant section
of the $\sym (4,\mathbb{R})$ bundle defined over $\mathrm{AdS_4}$.
This bundle is flat since the vanishing of the $\sym (4,\mathbb{R})$
curvature is nothing else but the Maurer-Cartan equation of $\sym
(4,\mathbb{R})$ and hence corresponds to the structural equations of
the $\mathrm{AdS_4}$ manifold. We are therefore guaranteed that
there exists a basis of four linearly independent sections of such a
bundle, namely four linearly independent solutions of
eq.(\ref{d4Killing}) which we can normalize as follows:
\begin{equation}
  \overline{\chi}_x \, \gamma_5 \, \chi_y \, = \, \epsilon_{xy}\,.
\label{normakillo4}
\end{equation}
The 1--forms on $\mathrm{AdS_4}$ are defined in terms of
$\mathrm{L_B}$ as follows:
\begin{equation}
- \, \ft 14 \, B^{ab} \, \gamma_{ab} \, - \,2\, e \, \gamma_a \,
\gamma_5 \,
 B^a \, = \, \Delta_B \, = \, \mathrm{L^{-1}_B} \, d \mathrm{L_B}\,.
\label{salamecanino}
\end{equation}
It follows that the inverse matrix $\mathrm{L^{-1}_B}$ satisfies the
equation:
\begin{equation}
  \left( d \, + \, \Delta_B \right) \, \mathrm{L^{-1}_B} \, = \, 0
\label{euchessina}
\end{equation}
Regarding the first index $y$ of the matrix $\left(
\mathrm{L^{-1}_B}\right)^y{}_x$ as the spinor index acted on by the
connection $\Delta_B$ and the second index $x$ as the labeling
enumerating the Killing spinors, eq.(\ref{euchessina}) is identical
with eq.(\ref{d4Killing}) and hence we have explicitly constructed
its four independent solutions. In order to achieve the desired
normalization (\ref{normakillo4}) it suffices to multiply by a phase
factor $\exp \left[{-\rm i} \, \ft 14 \pi \right]$, namely it
suffices to set:
\begin{equation}
  \chi^y_{(x)} \, = \, \exp \left[-{\rm i} \, \ft 14 \pi
\right] \, \left( \mathrm{L^{-1}_B}\right)^y{}_x \label{killini}
\end{equation}
In this way the four Killing spinors fulfill the Majorana condition,
having chosen a representation of the $D=4$ Clifford algebra in
which $\mathcal{C}=i\,\gamma_0$ (see Appendix \ref{d4spinorbasis}
for conventions on spinors). Furthermore since $\mathrm{L^{-1}_B}$
is symplectic it satisfies the defining relation
\begin{equation}
   \mathrm{L^{-1}_B} \, \mathcal{C} \,
  \gamma_5 \, \mathrm{L_B} \, = \,  \mathcal{C} \,
  \gamma_5 \,
\label{defirelazia}
\end{equation}
which implies (\ref{normakillo4}).
\subsection{Explicit construction of  $\mathbb{P}^3$ geometry}
The complex three-fold $\mathbb{P}^3$ is K\"ahler. Indeed the
existence of the K\"ahler $2$-form is one of the essential items in
constructing the solution ansatz.
\par
Let us begin by discussing all the relevant geometric structures of
$\mathbb{P}^3$. We need now to construct the explicit form of the
internal manifold geometry, in particular the spin connection, the
vielbein and the K\"ahler $2$-form. This is fairly easy, since
$\mathbb{P}^3$ is a coset manifold:
\begin{equation}
 \mathbb{P}^3 \, = \, \frac{\mathrm{SU(4)}}{\mathrm{SU(3) \times U(1)}}
\label{Ptre}
\end{equation}
so that everything is defined in terms of structure constants of the
$\su(4)$ Lie algebra. The quickest way to introduce these structure
constants and their chosen normalization is by writing the
Maurer--Cartan equations. We do this introducing already the
splitting:
\begin{equation}
  \su(4) \, = \, \mathbb{H} \, \oplus \, \mathbb{K}
\label{splittato}
\end{equation}
between the subalgebra $\mathbb{H} \, \equiv \, \su(3) \times
\uu(1)$ and the complementary orthogonal subspace $\mathbb{K}$ which
is tangent to the coset manifold. Hence we name $H^i \, (i=1,\dots,
9)$ a basis of $1$-form generators of  $\mathbb{H}$ and $K^\alpha \,
(\alpha =1,\dots, 6)$ a basis of $1$-form generators of
$\mathbb{K}$. With these notation the Maurer--Cartan equations
defining the structure constants of $\su(4)$ have the following
form:
\begin{eqnarray}
 dK^\alpha + \mathcal{B}^{\alpha\beta} \, \wedge \, K^\gamma \,
 \delta_{\beta \gamma} & = & 0 \nonumber\\
d\mathcal{B}^{\alpha \beta } \, + \, \mathcal{B}^{\alpha \gamma
} \, \wedge \, \mathcal{B}^{\delta \beta } \, \delta_{\gamma\delta} \, - \,\mathcal{X}^{\alpha \beta }_{\phantom{\alpha \beta}\gamma\delta} \,
K^\gamma \, \wedge \, K^\delta  &
= & 0
\label{maureCP3}
\end{eqnarray}
where:
\begin{enumerate}
  \item { the antisymmetric $1$-form valued matrix $B^{\alpha \beta }$ is
parametrized by the $9$ generators of the $\uu(3)$ subalgebra of
$\so(6)$ in the following way:
\begin{equation}
\mathcal{B}^{\alpha \beta } \, = \,  \left(
\begin{array}{llllll}
 0 & H^9 & -H^8 & H^1+H^2 & H^6 & -H^5 \\
 -H^9 & 0 & H^7 & H^6 & H^1+H^3 & H^4 \\
 H^8 & -H^7 & 0 & -H^5 & H^4 & H^2+H^3 \\
 -H^1-H^2 & -H^6 & H^5 & 0 & H^9 & -H^8 \\
 -H^6 & -H^1-H^3 & -H^4 & -H^9 & 0 & H^7 \\
 H^5 & -H^4 & -H^2-H^3 & H^8 & -H^7 & 0
\end{array}
\right)
\label{spinconP3}
\end{equation}}
  \item {the symbol $ \mathcal{X}^{\alpha \beta }_{\phantom{\alpha \beta}\gamma\delta}
  $ denotes the following constant, 4-index tensor:
  \begin{equation}
  \mathcal{X}^{\alpha \beta }_{\phantom{\alpha \beta}\gamma\delta} \,
  \equiv \,   \left( \delta^{\alpha \beta }_{\gamma \delta } \, +
  \, \mathcal{K}^{\alpha \beta } \, \mathcal{K}^{\gamma \delta } \, +
  \, \mathcal{K}^{\alpha}_{\phantom{\alpha } \gamma} \, \mathcal{K}^{\beta}_{\phantom{\beta } \delta} \,\right)
\label{Qtensor}
\end{equation}}
\item{
the symbol $\mathcal{K}^{\alpha \beta }$ denotes the entries of the
following antisymmetric matrix:
\begin{equation}
  \mathcal{{K}} \, = \, \left(
\begin{array}{llllll}
 0 & 0 & 0 & -1 & 0 & 0 \\
 0 & 0 & 0 & 0 & -1 & 0 \\
 0 & 0 & 0 & 0 & 0 & -1 \\
 1 & 0 & 0 & 0 & 0 & 0 \\
 0 & 1 & 0 & 0 & 0 & 0 \\
 0 & 0 & 1 & 0 & 0 & 0
\end{array}
\right)
\label{Khat}
\end{equation}}
\end{enumerate}
The Maurer Cartan equations (\ref{maureCP3}) can be reinterpreted as
the structural equations of the $\mathbb{P}^3$ 6-dimensional
manifold. It suffices to identify the antisymmetric $1$-form valued
matrix $\mathcal{B}^{\alpha \beta }$ with the spin connection and
identify the vielbein $\mathcal{B}^\alpha$ with the coset generators
$K^\alpha$, modulo a scale factor $\lambda$
\begin{equation}
  \mathcal{B}^\alpha \, = \, \frac{1}{\lambda} \, K^\alpha
\label{intevielb}
\end{equation}
With these identifications  the first of eq.s(\ref{maureCP3})
becomes the vanishing torsion equation, while the second singles out
the Riemann tensor as proportional to the tensor
$\mathcal{X}^{\alpha \beta }_{\phantom{\alpha \beta}\gamma\delta}$
of eq.(\ref{Qtensor}). Indeed we can write:
\begin{eqnarray}
\mathcal{R}^{\alpha \beta } & = & d\mathcal{B}^{\alpha \beta } \, + \, \mathcal{B}^{\alpha \gamma
} \, \wedge \, \mathcal{B}^{\delta \beta } \, \delta_{\gamma\delta}
\nonumber\\
& = & \mathcal{R}^{\alpha \beta }_{\phantom{\alpha \beta}\gamma\delta}
\mathcal{B}^{\gamma} \, \wedge \, \mathcal{B}^{\delta}
\label{2curvaP3}
\end{eqnarray}
where:
\begin{equation}
  \mathcal{R}^{\alpha \beta }_{\phantom{\alpha \beta}\gamma\delta} \,
  = \, \lambda^2 \, \mathcal{X}^{\alpha \beta }_{\phantom{\alpha \beta}\gamma\delta}
\label{rimanone}
\end{equation}
\par
Using the above Riemann tensor we immediately retrieve the explicit form of the Ricci tensor:
\begin{equation}
  \mathrm{Ric}_{\alpha\beta} \, = \, 4 \, \lambda^2 \,\eta_{\alpha\beta}
\label{riccilambda}
\end{equation}
For later convenience in discussing the compactification ansatz it is
convenient to rename the scale factor as follows:
\begin{equation}
 \lambda \, = \, 2\, e
\label{valuelambda}
\end{equation}
In this way we obtain:
\begin{equation}
  \mathrm{Ric}_{\alpha\beta} \, = \, 16 \, e^2 \,\eta_{\alpha\beta}
\label{riccilambda2}
\end{equation}
which will be recognized as one of the field equations of type IIA
supergravity.
\par
Let us now come to the interpretation of the matrix $\mathcal{K}$.
This matrix is immediately identified as  encoding the intrinsic components
of of the K\"ahler $2$-form. Indeed $ \mathcal{{K}}$ is the unique
antisymmetric matrix which, within the fundamental $6$-dimensional representation of the
$\so(6) \sim \su(4)$ Lie algebra, commutes with the entire subalgebra $\uu(3) \, \subset \,
\su(4)$. Hence  $\mathcal{K} $ generates the $\mathrm{U(1)}$ subgroup of
$\mathrm{U(3)}$ and this guarantees that the  K\"ahler $2$-form will
be closed and coclosed as it should be.
Indeed it is sufficient to set:
\begin{equation}
 \widehat{\mathcal{K}}\, = \,   \mathcal{K}_{\alpha \beta }  \, \mathcal{B}^\alpha  \,
 \wedge \, \mathcal{B}^\beta
\label{idekahler}
\end{equation}
namely:
\begin{equation}
  \widehat{\mathcal{K}} \, = \,- \, 2 \,   \left(  \mathcal{B}^1 \, \wedge \, \mathcal{B}^4 \, + \, \mathcal{B}^2 \, \wedge
  \,  \mathcal{B}^5 \, + \, \mathcal{B}^3 \, \wedge \, \mathcal{B}^6 \right)
\label{Kappoform}
\end{equation}
and  we obtain that the $2$-form $\widehat{\mathcal{K}}$ is closed and coclosed:
\begin{equation}
  d\, \widehat{\mathcal{K}} \, = \, 0 \quad , \quad d^\star \widehat{\mathcal{K}} \, = \, 0
\label{chiuca&cochiusa}
\end{equation}
Let us also note that the antisymmetric matrix $\mathcal{K}$
satisfies the following identities:
\begin{eqnarray}
\mathcal{K}^2 & = & - \,{1}_{6 \times 6} \nonumber\\
8 \, \mathcal{K}_{\alpha \beta } & = & \epsilon _{\alpha \beta \gamma \delta \tau \sigma
} \mathcal{K}^{\gamma \delta } \, \mathcal{K}^{\tau \sigma}
\label{preperK}
\end{eqnarray}
Using the $\so(6)$ Clifford Algebra defined in appendix \ref{d7spinorbasis}
we define the following spinorial operators:
\begin{equation}
  \mathcal{W} \, = \, {\mathcal{K}}_{\alpha \beta } \, \tau^{\alpha \beta } \quad
  ; \quad \mathcal{P} \, = \, \mathcal{W}\, \tau_7
\label{operatorini}
\end{equation}
and we can verify that the matrix $\mathcal{P}$ satisfies the
following algebraic equations:
\begin{equation}
  \mathcal{P}^2 +4\, \mathcal{P} -12 \, \times \, \mathbf{1} \, = \, 0
\label{agequadiP}
\end{equation}
whose roots are $2$ and $-6$. Indeed in the chosen $\tau$-matrix
basis the matrix $\mathcal{P}$ is diagonal with the following explicit form:
\begin{equation}
  \mathcal{P} \, = \, \left(
\begin{array}{llllllll}
 2 & 0 & 0 & 0 & 0 & 0 & 0 & 0 \\
 0 & 2 & 0 & 0 & 0 & 0 & 0 & 0 \\
 0 & 0 & 2 & 0 & 0 & 0 & 0 & 0 \\
 0 & 0 & 0 & 2 & 0 & 0 & 0 & 0 \\
 0 & 0 & 0 & 0 & 2 & 0 & 0 & 0 \\
 0 & 0 & 0 & 0 & 0 & 2 & 0 & 0 \\
 0 & 0 & 0 & 0 & 0 & 0 & -6 & 0 \\
 0 & 0 & 0 & 0 & 0 & 0 & 0 & -6
\end{array}
\right)
\label{ExpformadiP}
\end{equation}
Let us also introduce the following matrix valued $1$-form:
\begin{equation}
  \mathcal{Q} \, \equiv \, \left(\ft 32 \, \mathbf{1}  \, + \, \ft 14 \,
  \mathcal{P}  \right)\, \tau_\alpha \, \mathcal{B}^\alpha
\label{Qforma}
\end{equation}
whose explicit form in the chosen basis is the following one:
\begin{equation}
\mathcal{Q} \, = \,   \left(
\begin{array}{llllllll}
 0 & 2 \mathcal{B}^3  & -2 \mathcal{B}^2  & 0 & -2 \mathcal{B}^6  & 2 \mathcal{B}^5  & -2 \mathcal{B}^4  & 2 \mathcal{B}^1  \\
 -2 \mathcal{B}^3  & 0 & 2 \mathcal{B}^1  & 2 \mathcal{B}^6  & 0 & -2 \mathcal{B}^4  & -2 \mathcal{B}^5  & 2 \mathcal{B}^2  \\
 2 \mathcal{B}^2  & -2 \mathcal{B}^1  & 0 & -2 \mathcal{B}^5  & 2 \mathcal{B}^4  & 0 & -2 \mathcal{B}^6  & 2 \mathcal{B}^3  \\
 0 & -2 \mathcal{B}^6  & 2 \mathcal{B}^5  & 0 & -2 \mathcal{B}^3  & 2 \mathcal{B}^2  & 2 \mathcal{B}^1  & 2 \mathcal{B}^4  \\
 2 \mathcal{B}^6  & 0 & -2 \mathcal{B}^4  & 2 \mathcal{B}^3  & 0 & -2 \mathcal{B}^1  & 2 \mathcal{B}^2  & 2 \mathcal{B}^5  \\
 -2 \mathcal{B}^5  & 2 \mathcal{B}^4  & 0 & -2 \mathcal{B}^2  & 2 \mathcal{B}^1  & 0 & 2 \mathcal{B}^3  & 2 \mathcal{B}^6  \\
 0 & 0 & 0 & 0 & 0 & 0 & 0 & 0 \\
 0 & 0 & 0 & 0 & 0 & 0 & 0 & 0
\end{array}
\right)
\label{Qexplicit}
\end{equation}
and let us consider the following Killing spinor equation:
\begin{equation}
  \mathcal{D} \, \eta \, + \, e \, \mathcal{Q} \, \eta \, = \, 0
\label{killospinoequa}
\end{equation}
where, by definition:
\begin{equation}
  \mathcal{D} \, = \, d \, - \, \ft 14 \, \mathcal{B}^{\alpha\beta } \, \tau_{\alpha \beta
  }
\label{so6covderi}
\end{equation}
denotes the $\so(6)$ covariant differential of spinors defined over
the $\mathbb{P}^3$ manifold. The connection $\mathcal{Q}$ is closed
with respect to the spin connection
\begin{equation}
  \Omega \, = \, - \, \ft 14 \, \mathcal{B}^{\alpha\beta } \, \tau_{\alpha \beta
  }
\label{spinaconnaU3}
\end{equation}
since we have:
\begin{equation}
  \mathcal{D} \, \mathcal{Q} \, \equiv \, d\mathcal{Q} \, + \, e^2 \, \Omega \, \wedge \, \mathcal{Q} \,
  +
  \, \mathcal{Q} \, \wedge \, \Omega \, = \, 0
\label{closureQ}
\end{equation}
as it can be explicitly checked. The above result follows because the
matrix $\mathcal{K}_{\alpha \beta }$ commutes with all the generators of
$\uu(3)$.
In view of eq.(\ref{closureQ}) the integrability of the Killing
(\ref{killospinoequa}) becomes the following one:
\begin{equation}
 \mathrm{ Hol} \, \eta \, = \, 0
\label{integracondo}
\end{equation}
where we have defined the holonomy $2$-form:
\begin{equation}
  \mathrm{ Hol} \,\equiv \,  \left( \mathcal{D}^2 \, + \,e^2 \,  \mathcal{Q} \, \wedge \, \mathcal{Q}\right)
  \, = \, \left( - \, \ft 14 \, \mathcal{R}^{\alpha \beta } \, \tau_{\alpha \beta
  } \, + \, e^2 \, \mathcal{Q} \, \wedge \, \mathcal{Q}\right)
\label{Holodefi}
\end{equation}
and $\mathcal{R}^{\alpha \beta }$ denotes the curvature $2$-form
(\ref{2curvaP3}). Explicit evaluation of the holonomy $2$-form
yields the following result.
\begin{equation}
  \mathrm{Hol} \, = \, e^2 \, \left(
\begin{array}{llllllll}
 0 & 0 & 0 & 0 & 0 & 0 & 8 [\mathcal{B}^2 \wedge \mathcal{B}^6 -\mathcal{B}^3 \wedge \mathcal{B}^5 ] &
 8 \mathcal{B}^5 \wedge \mathcal{B}^6 -8 \mathcal{B}^2 \wedge \mathcal{B}^3  \\
 0 & 0 & 0 & 0 & 0 & 0 & 8 \mathcal{B}^3 \wedge \mathcal{B}^4 -8 \mathcal{B}^1 \wedge \mathcal{B}^6
 & 8 [\mathcal{B}^1 \wedge \mathcal{B}^3 -\mathcal{B}^4 \wedge \mathcal{B}^6 ] \\
 0 & 0 & 0 & 0 & 0 & 0 & 8 [\mathcal{B}^1 \wedge \mathcal{B}^5 -\mathcal{B}^2 \wedge \mathcal{B}^4 ]
 & 8 \mathcal{B}^4 \wedge \mathcal{B}^5 -8 \mathcal{B}^1 \wedge \mathcal{B}^2  \\
 0 & 0 & 0 & 0 & 0 & 0 & 8 [\mathcal{B}^2 \wedge \mathcal{B}^3 -\mathcal{B}^5 \wedge \mathcal{B}^6 ]
 & 8 [\mathcal{B}^2 \wedge \mathcal{B}^6 -\mathcal{B}^3 \wedge \mathcal{B}^5 ] \\
 0 & 0 & 0 & 0 & 0 & 0 & 8 \mathcal{B}^4 \wedge \mathcal{B}^6 -8 \mathcal{B}^1 \wedge \mathcal{B}^3
 & 8 \mathcal{B}^3 \wedge \mathcal{B}^4 -8 \mathcal{B}^1 \wedge \mathcal{B}^6  \\
 0 & 0 & 0 & 0 & 0 & 0 & 8 [\mathcal{B}^1 \wedge \mathcal{B}^2 -\mathcal{B}^4 \wedge \mathcal{B}^5 ]
 & 8 [\mathcal{B}^1 \wedge \mathcal{B}^5 -\mathcal{B}^2 \wedge \mathcal{B}^4 ] \\
 0 & 0 & 0 & 0 & 0 & 0 & 0 & -8 \, \widehat{\mathcal{K}} \\
 0 & 0 & 0 & 0 & 0 & 0 & 8 \, \widehat{\mathcal{K}} & 0
\end{array}
\right )
\label{Holoper}
\end{equation}
It is evident by inspection that the holonomy $2$-form vanishes on
the subspace of spinors that belong to the eigenspace of eigenvalue
$2$ of the operator $\mathcal{P}$. In the chosen basis this
eigenspace is spanned by all those spinors whose last two components
are zero and on such spinors the operator $\mathrm{Hol}$ vanishes.
\par
Let us now connect these geometric structures to the compactification
ansatz.
%%%%%%%%%%%%%%%%%%%%%%%%%%%%%%%
% BABBA
%%%%%%%%%%%%%%%%%%%%%%%%%%%%%%%%%%
%%%%%%%
\subsection{The compactification ansatz}
As usual we denote with latin indices those in the direction of
$4$-space and with Greek indices those in the direction of the
internal $6$-space. Let us also adopt the notation: $B^a$ for the
$\mathrm{AdS}_4$ vielbein just as $\mathcal{B}^{\alpha}$ is  the
vielbein of the K\"ahler three-fold described in the previous
section\footnote{This formulation is analogue to the one used in the
case of M-theory compactifications \cite{}}.  With these notations
the Kaluza-Klein ansatz is the following one:
\begin{eqnarray}
\mathcal{G}_{\underline{ab}} & = & \left \{\begin{array}{c}
 2 \, e \, \exp\left[-\varphi_0 \right] \, \mathcal{K}_{\alpha \beta } \\
  0 \quad \mbox{otherwise}
\end{array} \right. \nonumber\\
\mathcal{G}_{\underline{a_1a_2a_3a_4}} & = & \left \{\begin{array}{c}
- \, e \, \exp\left[-\varphi_0 \right] \, \epsilon_{a_1a_2a_3a_4} \\
 0 \quad \mbox{otherwise}
\end{array} \right. \nonumber\\
\mathcal{H}_{\underline{a_1a_2a_3}} & = & 0 \nonumber\\
\varphi & = & \varphi_0 \, = \, \mbox{constant}\,
\nonumber\\
V^a & = & B^{a} \nonumber\\
V^\alpha & = & \mathcal{B}^\alpha\nonumber\\
\omega^{ab} & = & B^{ab} \nonumber\\
\omega^{\alpha \beta } & = & \mathcal{B}^{\alpha \beta }
\label{Kkansatz}
\end{eqnarray}
where $B^a\, , \, B^{ab}$ respectively denote the vielbein and the spin connection of $\mathrm{AdS_4}$, satisfying the
following structural equations:
\begin{eqnarray}
0 & = & dB^a \, - \, B^{ab} \, \wedge \, B^c \, \eta_{bc} \nonumber\\
dB^{ab} \, - \, B^{ac} \, \wedge \, B^{db}  \, \eta_{cd}  & = & -
16 \, e^2 \,  \, B^a \, \wedge \, B^b \nonumber\\
& \Downarrow & \nonumber\\
\mbox{Ric}_{ab} & = & \, - \,  24\, e^2 \, \eta_{ab}
\label{ads4geo}
\end{eqnarray}
while $\mathcal{B}^\alpha$ and $\mathcal{B}^{\alpha \beta }$ are the
analogous data for the internal $\mathbb{P}^3$ manifold:
\begin{eqnarray}
0 & = & d\mathcal{B}^\alpha  \, - \, \mathcal{B}^{\alpha \beta } \, \wedge \, \mathcal{B}^\gamma  \, \eta_{ \beta\gamma } \nonumber\\
d\mathcal{B}^{\alpha \beta } \, - \, \mathcal{B}^{\alpha \gamma } \, \wedge \, \mathcal{B}^{\delta \beta }  \, \eta_{\gamma \delta }  & = & -
R^{\alpha \beta }_{\,\,\,\ \gamma \delta }\, \mathcal{B}^\gamma  \, \wedge \, \mathcal{B}^\delta  \nonumber\\
& \Downarrow & \nonumber\\
\mbox{Ric}_{\alpha \beta } & = &  16\, e^2 \, \eta_{\alpha \beta }
\label{HKgeo}
\end{eqnarray}
whose geometry we described in the previous section.
\par
 With these normalizations we can check that the dilaton equation (\ref{Rreq01})  and
  the Einstein equation (\ref{Einsteinus}), are satisfied upon insertion of the
above Kaluza Klein ansatz.
All the other equations are satisfied thanks to the fact that the K\"ahler form $\widehat{\mathcal{K}}$ is closed and coclosed:
eq.(\ref{chiuca&cochiusa})
\subsection{Killing spinors on $\mathbb{P}^3$}
The next task we are faced with is to determine the equation for the
Killing spinors on the chosen background, which by construction is a
solution of supergravity equations.
\par
Following a standard procedure we recall that the vacuum has been
defined by choosing certain values for the bosonic fields and setting
all the fermionic ones equal to zero:
\begin{eqnarray}
\psi_{L/R|\underline{\mu}} & = & 0 \nonumber\\
\chi_{L/R} & = & 0\nonumber\\
\rho_{L/R|\underline{ab}} & = & 0
\label{zerofermioni}
\end{eqnarray}
The equation for the Killing spinors will be obtained by imposing
that the parameter of supersymmetry preserves  the vanishing values of the fermionic
fields once the specific values of the bosonic ones is substituted into the
expression for the susy rules, namely into the rheonomic
parametrizations.
\par
To implement these conditions we begin by choosing a well adapted
basis for the $d=11$ gamma matrices. This is done by setting:
\begin{equation}
  \Gamma^{\underline{a}} \, = \, \left \{\begin{array}{ccc}
    \Gamma^a & = &  \, \gamma^a \, \otimes \,  \mathbf{1} \\
    \Gamma^\alpha & = & \gamma^5 \, \otimes \, \tau^\alpha \\
    \Gamma^{11} & = & {\rm i} \, \gamma^5 \, \otimes \, \tau^7 \
  \end{array} \, \right.
\label{productgamma}
\end{equation}
Next we consider the tensors and the matrices introduced in eq.s
(\ref{Mntensors},\ref{Mmatrapm},\ref{pongo},\ref{pongo2}). In the
chosen background we find:
\begin{eqnarray}
 \mathcal{M}_{\alpha \beta } &=& \ft 14 \, e \,
\mathcal{K}_{\alpha \beta }\,\,\,;\,\,\,\, \mathcal{M}_{abcd}  = \,
\ft 1{16} \, e \, \epsilon_{abcd}\nonumber\\
\mathcal{N}_0 & = & 0 \,\,\,;\,\,\,\, \mathcal{N}_{\alpha \beta }
=\ft 12 \, e \, \mathcal{K}_{\alpha \beta }\,\,\,;\,\,\,\,
\mathcal{N}_{abcd}  =  \, - \,\ft 1{24} \, e \, \epsilon_{abcd}\,,
\label{tensorisuvuoto}
\end{eqnarray}
all the other components of the above matrices being zero. Hence in
terms of the operators introduced in the previous section we find:
\begin{eqnarray}
\mathcal{M}_\pm & = & {\rm i} \, e \, \left(\mp \ft 14 \, \mathbf{1} \, \otimes \, \mathcal{W}  \, - \, \ft 32 \,{\rm i} \gamma_5 \,
\otimes \, \mathbf{1}\right)\nonumber\\
\mathcal{N}^{(even)}_\pm & = &  e \, \left(\ft 12 \, \mathbf{1} \, \otimes \, \mathcal{W}  \, \mp \, {\rm i} \gamma_5 \,
\otimes \, \mathbf{1}\right)\nonumber\\
\mathcal{N}^{(odd)}_\pm & = &  0
\label{valoritensori}
\end{eqnarray}
It is now convenient to rewrite the Killing spinor condition in a non
chiral basis introducing a supersymmetry parameter of the following
form:
\begin{equation}
  \epsilon \, = \, \epsilon _L \, + \, \epsilon _R
\label{nonchiral}
\end{equation}
In this basis the matrices $\mathcal{M}$ and $\mathcal{N}^{(even)}$
read
\begin{eqnarray}
\mathcal{M}&=&\mathcal{M}_+\,\frac{1}{2}\,(\bfone+\Gamma^{11})+\mathcal{M}_-\,\frac{1}{2}\,(\bfone-\Gamma^{11})
=-\frac{i}{8}\,e^{\varphi}\,G_{\underline{ab}}\,\Gamma^{\underline{ab}}\,\Gamma^{11}-
\frac{i}{16}\,e^{\varphi}\,G_{\underline{abcd}}\,\Gamma^{\underline{abcd}}=\nonumber\\&=&
\frac{e}{4}\,\gamma_5\otimes (\mathcal{W}\tau_7+6\,\bfone)\,,\label{calm}\\
\mathcal{N}^{(even)}&=&\mathcal{N}^{(even)}_+\,\frac{1}{2}\,(\bfone+\Gamma^{11})+\mathcal{N}^{(even)}_-\,\frac{1}{2}\,(\bfone-\Gamma^{11})=\frac{1}{4}\,e^{\varphi}\,G_{\underline{ab}}\,\Gamma^{\underline{ab}}+
\frac{1}{24}\,e^{\varphi}\,G_{\underline{abcd}}\,\Gamma^{\underline{abcd}}=\nonumber\\&=&
\frac{e}{2}\,\bfone\otimes (\mathcal{W}+2\tau_7)\,.\label{caln}
\end{eqnarray}
 Upon
use of this parameter the Killing spinor equation coming from the
gravitino rheonomic parametrization (\ref{rhoparaSF}) takes the
following form:
\begin{equation}
  \mathcal{D} \, \epsilon \, = - \mathcal{M}
  \, \Gamma_{\underline{a}} \, V^{\underline{a}} \,\epsilon \,,
\label{gravino}
\end{equation}
while the Killing spinor equation coming from the dilatino rheonomic parametrization is as follows:
\begin{equation}
  0 \, =\mathcal{N}^{(even)} \, \epsilon\,.
\label{ditalino}
\end{equation}
Let us now insert these results into the Killing spinor equations and
let us take a tensor product representation for the Killing spinor:
\begin{equation}
  \epsilon \, = \, \varepsilon \, \otimes \, \eta
\label{tensorerepre}
\end{equation}
where $\varepsilon$ is a $4$-component $d=4$ spinor and $\eta$ is an
$8$-component $d=6$ spinor.
\par
With these inputs equation (\ref{gravino}) becomes:
\begin{eqnarray}
0 & = & \mathcal{D}_{[4]}\varepsilon \, \otimes \, \eta \, - \,  e \, \gamma_a\,\gamma_5 \, B^a \varepsilon \otimes \,
\left(\ft 32 \, + \, \ft 14 \, \mathcal{P} \right)\, \eta  \nonumber\\
 & \null &  \, + \, \varepsilon \, \otimes \left[\mathcal{D}_{[6]} \,
 + \, e \, \left(\ft 32 \, + \, \ft 14 \, \mathcal{P} \right) \,
 \tau_\alpha \, \mathcal{B}^\alpha \right] \,\eta
\label{gravino2}
\end{eqnarray}
while eq.(\ref{ditalino}) takes the form:
\begin{equation}
0 \, = \,  \varepsilon \, \otimes \, \left( \ft 12 \, \mathcal{W} \, + \tau_7 \right) \,
  \eta
\label{ditalino2}
\end{equation}
Let us now recall that equation (\ref{killospinoequa}) is integrable
on the eigenspace of eigenvalue $2$ of the $\mathcal{P}$-operator.
Then equation (\ref{gravino2}) is satisfied if:
\begin{eqnarray}
\left( \mathcal{D}_{[4]} \,  - 2\,  e \, \gamma_a\,\gamma_5 \, B^a \right) \varepsilon & = &
0\nonumber\\
\mathcal{P}\, \eta & = & 2 \, \eta \nonumber\\
\left( \mathcal{D}_{[6]} \,
 + \, e \, \mathcal{Q}   \right) \,\eta & = & 0
\label{croma}
\end{eqnarray}
The first of the above equation is the correct equation for Killing
spinors in $\mathrm{AdS_4}$. It emerges if the eigenvalue of
$\mathcal{P}$ is $2$. The second and the third are the already
studied integrable equation for six Killing spinors out of eight. It
should now be that the dilatino equation (\ref{ditalino2}) is
satisfied on the eigenspace of eigenvalue $2$, which is indeed the
case:
\begin{equation}
  \mathcal{P}\, \eta \, = \, 2  \,\eta \, \Rightarrow \, \left( \ft 12 \, \mathcal{W} \, + \tau_7 \right) \,
  \eta \, = \, 0
\label{rimasuglio}
\end{equation}
\subsection{Gauge completion in  mini superspace}
As a necessary ingredient of our construction let $\eta_A$ ($A=1,\dots,6$) denote a complete and orthonormal  basis of solutions
the internal Killing spinor equation, namely:
\begin{eqnarray}
\mathcal{P}\, \eta_A & = & 2 \, \eta_A \nonumber\\
\left( \mathcal{D}_{[6]} \,
 + \, e \, \mathcal{Q}   \right) \,\eta_A & = & 0 \nonumber\\
 \eta^T_A \, \eta_B & = & \delta_{AB} \quad ; \quad A,B \, = \,A=1,\dots,6
\label{basispinotti}
\end{eqnarray}
On the other hand let $\chi_x$ denote a basis of solutions of the
Killing spinor equation on $AdS_4$-space, namely (\ref{d4Killing}) ,
normalized as in eq.(\ref{normakillo4}). Furthermore let us recall
the matrix $K$ defining the intrinsic components of the K\" ahler
$2$-form.
\par
In terms of these objects we can satisfy the rheonomic
parametrizations of the $1$-forms spanning the $d=10$ superPoincar\'e
subalgebra of the FDA with the following position:\footnote{With respect to the results obtained in \cite{Fre:2007xy}
for the mini superspace extension of M-theory configuration
everything is identical in eq.s(\ref{graviansaz}-\ref{omabAnsaz})
except the obvious reduction of the index range of
($\alpha,\beta,\dots$) from $7$ to $6$-values. The only difference
is in eq.(\ref{omalbetAnsaz}) where the last contribution
proportional to the K\"ahler form is an essential novelty of this
new type of compactification.}

\begin{eqnarray}
\Psi & = & \chi_x \, \otimes \, \eta_A \, \Phi^{x|A} \label{graviansaz}\\
V^a & = & B^a \, - \, \ft {1}{8e} \, \overline{\chi}_x \, \gamma^a \,
\chi_y \, \Delta^{xy} \label{VaAnsaz}\\
V^\alpha & = & \mathcal{B}^\alpha \, - \, \ft {1}{8} \, \eta_A^T \, \tau^\alpha \,
\eta_B \, \mathcal{A}^{AB} \label{ValpAnsaz}\\
\omega ^{ab} & = & B^{ab} \, + \, \ft {1}{2} \, \overline{\chi}_x \, \gamma^{ab}
\,\gamma_5 \,
\chi_y \, \Delta^{xy} \label{omabAnsaz}\\
\omega^{\alpha \beta}  & = & \mathcal{B}^{\alpha \beta } \, + \, \ft {e}{4} \, \eta_A^T \, \tau^{\alpha\beta } \,
\eta_B \, \mathcal{A}^{AB}  \, - \, \ft e4 \mathcal{K}^{\alpha \beta } \, \mathcal{K}_{AB} \, \mathcal{A}^{AB}\label{omalbetAnsaz}
\end{eqnarray}
\par
The proof that the above ansatz satisfies the rheonomic
parametrizations is by direct evaluation upon use of the following
crucial spinor identities.
\par
Let us define
\begin{equation}
  \mathcal{U} \, = \, \left( \ft 32 \, \mathbf{1} \, + \, \ft 14 \,
  \mathcal{P} \right)\,.
\label{prillo1}
\end{equation}
We can verify that:
\begin{equation}
  \left( \eta_A \, \tau^\alpha \, \mathcal{U} \, \tau^\alpha \,
  \eta_B \, - \, \eta_A \, \tau^{\alpha\beta} \,
  \eta_B \right) \, \mathcal{A}^{AB} \, = \,  \mathcal{K}^{\alpha \beta } \, \mathcal{K}_{AB}
  \, \mathcal{A}^{AB}\,.
\label{fortedeimarmi}
\end{equation}
Furthermore, naming:
\begin{eqnarray}
\Delta \mathcal{B}^\alpha & = &  - \, \ft {1}{8} \, \eta_A^T \, \tau^\alpha \,
\eta_B \, \mathcal{A}^{AB} \label{DalpAnsaz}\\
\Delta\omega^{\alpha \beta}  & = &  \ft {e}{4} \, \eta_A^T \, \tau^{\alpha\beta } \,
\eta_B \, \mathcal{A}^{AB}  \, - \, \ft e4 \mathcal{K}^{\alpha \beta } \, \mathcal{K}_{AB} \, \mathcal{A}^{AB}\label{DomalbetAnsaz}
\end{eqnarray}
we obtain:
\begin{equation}
  - \,\Delta\omega^{\alpha \beta} \, \wedge \, \Delta
  \mathcal{B}^\beta \, = \, \ft {e}{8} \, \eta_A^T \, \tau^\alpha \,
\eta_B \, \mathcal{A}^{AC} \, \wedge \, \mathcal{A}^{CB}
\label{ferdison}
\end{equation}
These identities together with the $d=4$ spinor identities
(\ref{firzusd4x1},\ref{firzusd4x2}) suffice to verify that the above
ansatz satisfies the required equations.
%%%%%%%%%%%%%%%%%%%
\subsection{Gauge completion of the $\mathbf{B}^{[2]}$ form}
The next task in order to write the explicit form of the pure spinor
sigma-model is the derivation of the explicit expression for the
$\mathbf{B}^{[2]}$ form. When this is done we will be able to write
the complete Green Schwarz action in explicit form.
\par
There is an ansatz for $\mathbf{B}^{[2]}$ which is the following one:
\begin{equation}
  \mathbf{B}^{[2]} \, = \, \alpha \, \, \overline{\chi}_x \, \chi_y \, \overline{\eta}_A \,
  \tau_7 \,
  \eta_B \, \Phi^x_A \, \wedge \, \Phi^y_B
\label{ronron}
\end{equation}
By explicit evaluation we verify that with
\begin{equation}
  \alpha \, = \, \frac{1}{4 \, e}
\label{ruppo}
\end{equation}
The rheonomic parametrization of the H-field strength is satisfied,
namely:
\begin{equation}
  d \mathbf{B}^{[2]} \, = \, - {\rm i} \, \overline{\psi} \,\wedge \, \Gamma_{\underline{a}} \, \Gamma_{11} \,
  \psi \, \wedge \,  V^{\underline{a}}
\label{b2eque}
\end{equation}
\subsection{Rewriting the mini-superspace gauge completion as MC forms on
the complete supercoset} Next, following the procedure introduced in
\cite{Fre':2006es}, we rewrite the mini-superspace extension of the
bosonic solution solely in terms of Maurer Cartan forms on the
supercoset (\ref{supermanifoldamente}). Let the graded matrix
$\mathbb{L} \, \in \, \mathrm{Osp(6|4)}$ be the coset representative
of the coset $\mathcal{M}^{10|24}$, such that the Maurer Cartan form
$\Sigma$ can be identified as:
\begin{equation}
  \Sigma = \mathbb{L}^{-1} \, d \mathbb{L}
\label{cosettusrepre2}
\end{equation}
Let us now factorize $\mathbb{L}$ as in \cite{Fre':2006es}:
\begin{equation}
  \mathbb{L} = \mathbb{L}_F \, \mathbb{L}_B
\label{factorL2}
\end{equation}
where $\mathbb{L}_F$ is a coset representative for the coset :
\begin{equation}
  \frac{\mathrm{Osp(6 | \, 4
})}{\mathrm{SO(6)} \times \mathrm{Sp(4,\mathbb{R})}} \, \ni \,
\mathbb{L}_F \label{LF2}
\end{equation}
just in eq.(\ref{LF2}) but $\mathbb{L}_B$ rather than being  the
$\mathrm{Osp(6|4)}$ embedding of a coset representative of just
$\mathrm{AdS_4}$, is the embedding of a coset representative of
$\mathrm{AdS_4} \times \mathbb{P}^3$, namely:
\begin{equation}
  \mathbb{L}_B \, = \, \left(\begin{array}{c|c}
    \mathrm{L_{\mathrm{AdS_4}}} & 0 \\
    \hline
    0 & \mathrm{L_{{\mathbb{P}}^3}} \
  \end{array} \right) \quad ; \quad
  \frac{\mathrm{Sp(4,\mathbb{R})}}{\mathrm{SO(1,3)}}\, \ni \,
  \mathrm{L_{\mathrm{AdS_4}}} \quad ; \quad
  \frac{\mathrm{SO(6)}}{\mathrm{U(3)}}\, \ni \,
  \mathrm{L_{{\mathbb{P}}^3}}
\label{salamefelino2}
\end{equation}
In this way we find:
\begin{equation}
  \Sigma = \mathbb{L}_B^{-1} \, \Sigma_F \, \mathbb{L}_B \, + \, \mathbb{L}_B^{-1}
  \, d \, \mathbb{L}_B
\label{ferrone2}
\end{equation}
Let us now write the explicit form of $\Sigma_F $, as in
\cite{Fre':2006es}:
\begin{equation}
  \Sigma_F =\left(\begin{array}{c|c}
    \Delta_F & \Phi_A \\
    \hline
    \,  4 \, {\rm i} \, e \, \overline{\Phi}_A \, \gamma_5 & - \, e \, \widetilde{\mathcal{A}}_{AB} \
  \end{array} \right)
\label{fermionform2}
\end{equation}
where $\Phi_A$ is a Majorana-spinor valued fermionic $1$-form and
where $\Delta_F$ is an $\sym(4,\mathbb{R})$ Lie algebra valued
$1$-form presented as a $4 \times 4$ matrix. Both $\Phi_A$ as
$\Delta_F$ and $\widetilde{\mathcal{A}}_{AB}$ depend only on the
fermionic $\theta$ coordinates and differentials.
\par
On the other hand we have:
\begin{equation}
  \mathbb{L}_B^{-1}
  \, d \, \mathbb{L}_B \, = \, \left(\begin{array}{c|c}
    \Delta_{\mathrm{AdS_4}} & 0 \\
    \hline
    0 & \mathcal{A}_{\mathbb{P}^3} \
  \end{array} \right)
\label{boseforma2}
\end{equation}
where the $\Delta_{\mathrm{AdS_4}}$ is also an $\sym(4,\mathbb{R})$
Lie algebra valued $1$-form presented as a $4 \times 4$ matrix, but
it depends only on the bosonic coordinates $x^\mu$ of the anti de
Sitter space $\mathrm{AdS_4}$. In the same way
$\mathcal{A}_{\mathbb{P}^3}$ is an $\su(4)$ Lie algebra element
presented as an $\so(6)$ antisymmetric matrix in $6$-dimensions. It
depends only on the bosonic coordinates $y^\alpha$ of the internal
$\mathbb{P}^3$ manifold. According to eq(\ref{lambamatra}) we can
write:
\begin{equation}
\Delta_{\mathrm{AdS_4}} \, = \,    -  \ft 14 \, B^{ab} \,
\gamma_{ab} \, - \,2\, e \, \gamma_a \, \gamma_5 \, B^a
\label{Bbwriting2}
\end{equation}
where $\left\{ B^{ab}\, ,\, B^a\right\} $ are respectively the
spin-connection and the vielbein of $\mathrm{AdS_4}$.
\par
Similarly, using the inversion formula (\ref{gonzalo}) presented in
appendix we can write:
\begin{equation}
 \mathcal{A}_{\mathbb{P}^3} \, = \left( - \, 2 \,  \mathcal{B}^{\alpha } \, {\bar
  \tau}_\alpha \, + \, \ft{1} {4 \, e} \,  \mathcal{B}^{\alpha\beta
} \, {\bar \tau}_{\alpha \beta } - \, \ft{1} {4 \, e} \,
\mathcal{B}^{\alpha\beta } \, {\mathcal{K}}_{\alpha \beta } \,
K\right) \label{cuzco}
\end{equation}
where  $\left\{ \mathcal{B}^{\alpha\beta}\, ,\,
\mathcal{B}^\alpha\right\}$ are the connection and vielbein of the
internal coset manifold $\mathbb{P}^3$.
\par
Relying once again on the inversion formulae discussed in the
appendix we conclude that we can rewrite eq.s (\ref{graviansaz} -
\ref{omalbetAnsaz}) as follows:
\begin{eqnarray}
\Psi^{x|A} & = &  \Phi^{x|A} \label{graviansaz2}\\
V^a & = & E^a  \label{VaAnsaz2}\\
V^\alpha & = & E^\alpha   \label{ValpAnsaz2}\\
\omega ^{ab} & = & E^{ab}   \label{omabAnsaz2}\\
\omega^{\alpha \beta}  & = & E^{\alpha \beta } \label{omalbetAnsaz2}
\end{eqnarray}
where the objects introduced above are the MC forms on the
supercoset (\ref{cosettone1024}) according to:
\begin{equation}
   \Sigma \, =\, \mathbb{L}^{-1} \, d \mathbb{L} \, = \, \left(\begin{array}{c|c}
    -  \ft 14 \, E^{ab} \, \gamma_{ab} \, - \,2\, e \, \gamma_a \, \gamma_5 \, E^a  & \Phi \\
    \hline
    \,  4 \, {\rm i} \, e \, \overline{\Phi} \, \gamma_5 & \, 2 \, e  E^{\alpha } \, {\bar
  \tau}_\alpha \, - \, \ft{1} {4 } \,  \mathcal{B}^{\alpha\beta
} \, {\bar \tau}_{\alpha \beta } + \, \ft{1} {4 } \, E^{\alpha\beta
} \, {\mathcal{K}}_{\alpha \beta } \, K \
  \end{array} \right)
\label{pattolina}
\end{equation}
Consequentely the gauge completion of the $\mathbf{B}^{[2]}$ form
becomes:
\begin{equation}
  \mathbf{B}^{[2]} \, = \, \frac{1}{4 \, e} \,  \overline{\Phi}\, \left( 1 \, \otimes \, \overline{\tau}_7 \right) \wedge \, \Phi
\label{ronron2}
\end{equation}
%%%%%%%%%%%%%%%%%%%%%%%%%%%%%%%%%%%%%%%
\section{Pure Spinors for ${\rm Osp}(6|4)$}
In the present section, we show that the number of independent pure spinor components
obtained by solving the pure spinor constraint in the present background matches correctly the
number of anticommuting $\theta$'s. This implies that, at least formally (since it must be proved
in detail) the number of bosonic and fermionic fields match leading to a conformal invariant theory.
However, as is known, this is not sufficient for having a conformal invariant theory since all loop
contributions to the Weyl anomaly should cancel. This can be guaranteed only by symmetry reasons and for the vanishing of one-loop contribution.
\par
Nevertheless, we study the pure spinor equations adapted to the present background and we
will see that the number of the independent components of the pure spinors is equal to 14
(since we have an interacting theory with RR fields we cannot distinguish between left- and right-movers).
We recall the form of the pure spinor constraints for type IIA theory
\begin{eqnarray}
\label{psA}
& \bar\lambda \Gamma_{\underline a} \lambda =0\,,  ~~~~~~~~
 \bar\lambda \Gamma_{\underline a} \Gamma^{11} \lambda \, V^{\underline a} =0\,, \\
& \bar\lambda \Gamma_{[\underline{ab}]}  \lambda \,
V^{\underline a} V^{\underline b} =0\,, ~~~~~~
 \bar\lambda \Gamma^{11} \lambda =0 \,.
\end{eqnarray}
where we have combined the 16-component spinors $\lambda_1$ and $\lambda_2$ into a
32-component Dirac spinor $\lambda$.
These equations are valid for any background and we have shown in \cite{Psconstra}
the number of independent components for the pure spinors matches the number of pure spinor
in the Berkovits' "background-independent" constraints. However, in the present setting
we can adapt the constraints to the specific background and in particular we
choose to embed the vielbein $V^{\underline a}$ using his equation of motion in the momentum
$\Pi^{\underline a}_{\pm} e^{\pm}$ and thus simplifying the constraints as follows
\begin{eqnarray}
\label{psB}
&
\bar\lambda \Gamma_{a} \lambda =0\,,  ~~~~~~~~a=1,\dots,4\,,
~~~~~~
 \bar\lambda \Gamma_{\alpha} \lambda =0\,,  ~~~~~~~~ \alpha =1, \dots,6\,, \\
&
\bar\lambda \Gamma_{\pm} \Gamma^{11} \lambda =0\,,  ~~~~~~
\bar\lambda \Gamma_{+-}  \lambda =0\,, ~~~~~~
\bar\lambda \Gamma^{11} \lambda = 0 \,.
\end{eqnarray}
For $\Gamma_{\pm}$ we use the combination $\Gamma_1 \pm \Gamma_3$.
\par
Now, we can insert the decomposition of $\lambda$ on the basis of
Killing spinors
\begin{eqnarray}
\lambda &=& \chi_x \otimes \eta_A \, \Lambda^{x|A}\label{pureans}
\end{eqnarray}
 where, as usual, $\chi_x$
are the $AdS_4$-Killing spinors and $\eta_A$ are the $\mathbb{CP}^3$
Killing spinors. The free parameters $\Lambda^{x|A}$ are the
components the pure spinors. Notice that the index $x$ runs over the
four independent $AdS$-Killing spinor basis and the index $A$ runs
over the six values of vector representation of $SO(6)$.Therefore,
we have in total 24 independent degrees of freedom to solve
(\ref{psB}). The number of equations is independent of the
backgorund, but the number of independent degrees of freedom is
reduced from 32 to 24 and therefore, we need to explore the
esistence of the solution.
\par
Using the decomposition of the Gamma matrices provided in (\ref{productgamma}) and the
normalizations of the Killing spinors $\chi_x C \gamma_5 \chi_y = \e_{xy}$ and
$\eta_A \eta_B = \delta_{AB}$ , equations (\ref{psB}) read
\begin{eqnarray}
\label{psC}
&
(\chi_x C \gamma_{a} \chi_y) \, \delta_{AB} \, \Lambda^{x|A} \Lambda^{y|B} =0\,, ~~~~~~
(\chi_x C \gamma_5 \chi_y) \,  \eta_{A} \tau^\alpha \eta_B \,  \Lambda^{x|A} \Lambda^{y|B} =0\,,  \\
&
(\chi_x C \gamma_5 \chi_y)\,  \eta_{A} \tau^7 \eta_B  \, \Lambda^{x|A} \Lambda^{y|B} =0\,,  \\
&
(\chi_x C \gamma_5 \gamma_{\pm} \chi_y)\,  \eta_{A} \tau^7 \eta_B  \, \Lambda^{x|A} \Lambda^{y|B} =0\,,
~~~~~~
(\chi_x C \gamma_{+-} \chi_y) \,  \delta_{AB} \,  \Lambda^{x|A} \Lambda^{y|B} =0\,.
\end{eqnarray}
where $C$ is charge conjugation matrix.
\par
To solve these equations is convenient to adopt a new basis. Since we already know the solution
in the basis when the spinor $\Lambda$ is decomposed as follows
\begin{equation}
  \lambda_1 \, = \, \phi_+ \, \otimes \, \zeta_1^+ \, + \, \phi_- \, \otimes \, \zeta^-_1 \,,  ~~~~
  \lambda_2 \, = \, \phi_+ \, \otimes \, \zeta_2^- \, + \, \phi_- \, \otimes \, \zeta^+_2
\label{tensoproducto}
\end{equation}
where:
\begin{equation}
  \begin{array}{ccccccc}
    \phi_+ & = & \left( \begin{array}{c}
      1 \\
      0 \\
    \end{array}\right)
     & ; & \phi_- & = & \left(\begin{array}{c}
       0 \\
       1 \\
\end{array} \right)  \\
    \zeta_A^+ & =  & \left(\begin{array}{c}
      0 \\
      \omega^+_A \
    \end{array}  \right) & ; & \zeta_A^- & =  & \left( \begin{array}{c}
      \omega_A^- \\
      0 \
    \end{array}\right)  \
  \end{array}
\label{blocchini}
\end{equation}
where $\omega^{\pm}_A$ are $8$-dimensional vectors.
In writing eq.s~(\ref{blocchini}) we have observed that the unique
component of $\phi_\pm$ can always be reabsorbed in the normalization
of $\omega_A^\pm$ and hence set to one. Thus, we have to express the entries of the
rectangular matrix $\Lambda^{x|A}$ in terms of $\omega^{\pm}_A$ ($A=1,2$) and this can be done
by combining $\lambda_{1}$ and $\lambda_2$ in a single 32-dimensional pure spinor and projecting
it on the basis formed by $\chi_x \otimes \eta_A$ (where we left $A$ running over 8 values) and we
get the relation
\begin{equation}\label{psE}
\Lambda^{x|A} =
\left(
\begin{array}{ccc}
 \omega^-_{2,1} & \dots  & \omega^-_{2,8}  \\
 - {\mathrm i}\, \omega^+_{1,1} & \dots  &  - {\mathrm i} \omega^+_{1,8}  \\
 - {\mathrm i}\, \omega^-_{1,1} & \dots  &  - {\mathrm i} \omega^-_{1,8}  \\
 \omega^+_{2,1} & \dots  & \omega^+_{2,8}  \\
\end{array}
\right)
\end{equation}
In order to reduce the number of components to the neccessary 24 ones, we will set the last
components $\omega^{\pm}_{A,7}$ and $\omega^{\pm}_{A,8}$ to zero. In order to check
if this is possible it is convenient first to exploit all gauge symmetries.
\par
We recall that $\lambda_A$ are solutions of the constraints if
the components $\omega^{\pm}_A$ are decomposed in the following way
\begin{eqnarray}
\omega^+_1 & = & \left(\varpi^\alpha \, , \, 0 \right)  \nonumber\\
\omega^-_2 & = & \left(\pi^\alpha \, , \, 0 \right)  \nonumber\\
\omega^-_1 & = & \left(a^{\alpha \beta \gamma }\, \chi_\beta \, \varpi_\gamma \, , \, \chi\, \cdot \, \varpi
\right)\nonumber\\
\omega^+_2 & = & \left(a^{\alpha \beta \gamma }\, \xi_\beta \, \pi_\gamma \, , \, \xi\, \cdot \, \pi
\right)
\label{soluzia}
\end{eqnarray}
in terms of $7$-component fields
$\varpi^\alpha\,, \pi^\alpha\,, \xi^\alpha\,, \chi^\alpha$
satisfying  the constraints
\begin{eqnarray}
  \varpi \, \cdot \, \varpi & = & 0 \label{purga1}\\
\pi \, \cdot \, \pi & = & 0 \label{purga2}\\
a^{\alpha \beta \gamma } \, \chi_\alpha \, \pi_\beta \, \varpi_\gamma
& = & 0 \label{purga3}\\
a^{\alpha \beta \gamma } \, \xi_\alpha \, \pi_\beta \, \varpi_\gamma
& = & 0\,. \label{purga4}
\end{eqnarray}
Here $a^{\alpha\beta\gamma}$ is the totally-antisymmetric invariant tensor for $\mathrm{G_2}$ group.
Notice that constraints (\ref{purga1})-(\ref{purga4}) are invariant under the gauge symmetry
\begin{equation}\label{psF}
\chi_\alpha \rightarrow \chi_\alpha + x_1 \pi_\alpha + x_2 \varpi_\alpha\,, ~~~~~~~
\xi_\alpha \rightarrow \xi_\alpha + x_3 \pi_\alpha + x_4 \varpi_\alpha\,, ~~~~
\end{equation}
On the other side, the decomposition (\ref{soluzia}) is not invariant under the symmetries
parameterized by $x_1$ and $x_4$. So, there are only two gauge symmetries generated by $x_2$
and $x_3$ which can be used to set some components of $\chi_\alpha$ and $\xi_\alpha$  to zero.
\par
In order to reduce the number of independent degrees of freedom from
32 to 24, we set $\varpi^7$ and $\pi^7$ to zero, this condition,
together with (\ref{purga1}) and (\ref{purga2}), implies that
$\omega^+_1$ and $\omega^-_2$ have respectively 5 and 5 independent
degrees of freedom. In addition, we impose the equations
\begin{eqnarray}\label{psG}
&\chi\, \cdot \, \varpi =0\,, ~~~~~
 a^{7 \, \beta \gamma }\, \chi_\beta \, \varpi_\gamma =0\,, \\
&
\xi\, \cdot \, \pi =0 \,, ~~~~~
a^{7\, \beta \gamma }\, \xi_\beta \, \pi_\gamma =0\,.
\end{eqnarray}
such that the 7$^{\rm th}$ and the 8$^{\rm th}$ components of
$\Lambda^{x|A}$ are zero. Together with constraints (\ref{purga3})
and (\ref{purga4}), they can be solved in terms of 3 components of
$\chi_\alpha$ and 3 components $\xi_\alpha$. This reduces the number
of unfixed components from 14 to 8. Using the gauge symmetries
(\ref{psF}), we can lower them to 6 unfixed components. Finally,
observe that there are two additional gauge symmetries generated by
the constraints $\pi^7=0$ and $\varpi^7=0$ which reduce the number
of unfixed parameters for $\chi_\alpha$ and $\xi_\alpha$ to 4. The
total counting of the pure spinor conditions, in the space of 24
components of the matrix $\Lambda^{x|A}$, is exactly 14 (5 for
$\varpi$, 5 for $\pi$, 2 for $\chi$ and 2 for $\xi$), which is the
correct number of degrees of freedom in order to cancel the total
central charge. Indeed, we have $10$ from the boson $x^{\underline
a}$, 24 for $\theta$'s and the bosons $\Lambda$ which are 14 cancel
the total charge.
\par
In addition, we can compute the number of the conjugate fields for
the $\theta$ and for $w$ and using the constraints and the gauge
symmetry it is easy to prerform the same computations as in
\cite{Psconstra}  to see that the number matches again.

\section{Action}\label{s4}
Following the notations of \cite{D'Auria:2008ny} the complete action
of Pure Spinor superstrings on Type IIA backgrounds is the sum of two
parts, the Green-Schwarz action plus the gauge-fixing action
containing the pure spinor sector:
\begin{equation}
  \mathcal{A}^{IIA}_{PS} \, = \,  \int \mathcal{L}_{GS} \, +
  \,  \int \mathcal{L}^{IIA}_{gf}\,,
\label{lulla1}
\end{equation}
The GS action is written as follows
\begin{eqnarray}
  \mathcal{L}_{GS} = \left( \Pi^{\underline{a}}_+ \,
  V^{\underline{b}} \, \eta_{\underline{ab}} \, \wedge \, e^+ \, - \,
  \Pi^{\underline{a}}_- \,
  V^{\underline{b}} \, \eta_{\underline{ab}} \, \wedge \, e^-
  %\,\right.\nonumber\\
 %&& \left.
  + \, \ft 12 \Pi^{\underline{a}}_i\, \Pi^{\underline{b}}_j
  \, \eta^{ij}\, \eta_{\underline{ab}} \, e^+ \, \wedge \, e^-  \right )\,
  + \ft 12  \, \mathbf{B}^{[2]}\,.
\label{2akinact}
\end{eqnarray}
where $\Pi^{\underline a}_\pm$ are auxiliary fields whose field
equations identify them with the pull-back of the target-space
vielbein $V^{\underline a}$ on the worldsheet respectively along the
zweibein $e^+$ and $e^-$. $\eta_{ij}$ and $\eta_{\underline ab}$ are
the Minkowskian flat metrics respectively on the worldsheet and on
the 10d target space. The variation in the zweibein yields the
Virasoro constraints. The background geometry of the worldsheet
encoded in the reference frame $e^\pm$ is treated classically
\cite{Berkovits:2007wz,Hoogeveen:2007tu}.\par The gauge-fixing terms
of the string-action is written in \cite{D'Auria:2008ny} as:
\begin{eqnarray}\label{expA}
\mathcal{L}_{gf}^{\mathrm{IIA}} & = &  \overline{\mathbf d}_+ \,
\psi_R \, \wedge \, e^+ + \overline{\mathbf d}_-  \, \psi_L \,
\wedge \, e^-   +  \frac{\rm i}{2}
\overline{\mathbf d}_+  \,  \mathcal{M}_-  \, {\mathbf  d}_- \nonumber \\
&+& \overline{w}_+  \mathcal{D}\lambda_R \, \wedge \, e^+
+ \overline{w}_- \, \mathcal{D}\lambda_L \, \wedge \, e^-  \nonumber \\
&-& \frac{\rm i}{2} \,  \overline{w}_+  \left( \mathcal{S}_{R}
\mathcal{M}_-  \right) {\mathbf  d}_- + \frac{\rm i}{2} \,
\overline{\mathbf d}_+  \left( \mathcal{S}_{L}  \mathcal{M}_-
\right) {w}_- \nonumber\\
&-& \frac{\rm i}{2} \,  \overline{w}_+  \left(
\mathcal{S}_{R} \mathcal{S}_{L}  \mathcal{M}_-  \right) {w}_- +
\frac{\rm i}{2} \overline{w}_+ {\mathcal {M}}_- \{{\mathcal S}_L, {\mathcal S}_R\} w_-
\,.
\end{eqnarray}
The operators $\mathcal{S}_{L/R}$ represent the components of the
BRST operator $\mathcal{S}$ which are parametrized by the left/right
components of the pure spinor $\lambda$. The subscript $\pm$ on the
spinor matrices refer to their action on fermions with  left/right
chirality respectively. The last term is generated by the
non-vanishing the ${\mathcal S}_L {\mathcal S}_R$-piece of the
action in \cite{D'Auria:2008ny}. With reference to
\cite{D'Auria:2008ny}, we note that on the considered background the
operator $\hat{\mathcal S}_{L/R}$ coincide with ${\mathcal S}_{L/R}$
since ${\mathcal H}^{abc}$ field strength vanishes in this case.

The bosonic background corresponding to the $\mathrm{AdS_4}\times
\mathbb{P}^3$ solution of Type IIA theory is characterized by the
 values of the background fields displayed  in eq.(\ref{Kkansatz}).
The spinor matrices $\mathcal{M}$ and $\mathcal{N}^{(even)}$,
encoding the RR field-strengths, are given in eqs. (\ref{calm}),
(\ref{caln}) respectively. The matrix ${\mathcal M}$ in the present
background is constant and, therefore we can eliminate the auxiliary
fields ${\mathbf d}_\pm$ and write the complete quadratic part of
the action in terms of the MC forms. We start from the first two
lines of (\ref{expA})
\begin{equation}\label{acA}
 {\cal L}^{IIA}_{gf, 2} = \overline{\mathbf d}_+ \, \psi_R \, \wedge \, e^+ +
\overline{\mathbf d}_-  \, \psi_L \, \wedge \, e^-   +  \frac{\rm i}{2}
\overline{\mathbf d}_+  \,  \mathcal{M}_-  \, {\mathbf  d}_- \, e^+ \wedge e^-
\,.
\end{equation}

 We use the the decomposition of the gravitinos
 $$\Psi = \psi_+ e^+ + \psi_- e^- =
 \chi_x \otimes \eta_A (\Phi^{x A}_+ \, e^+ +\Phi^{x A}_- \, e^-)\,,$$
 where the 1-form is pull-back onto the worldsheet, then (\ref{acA}) yields
\begin{equation}\label{acB}
 {\cal L}^{IIA}_{gf, 2} = \Big(  -  {\mathbf d}^T_+ \frac{C (1 - \Gamma_{11})}{2}\, \psi_-  +
{\mathbf d}^T_-  \, \frac{C (1 + \Gamma_{11})}{2}\, \psi_+
+  \frac{\rm i}{2}
{\mathbf d}^T_+  \,  C\, \mathcal{  M}_-  \, {\mathbf  d}_-   \Big) \, e^+ \wedge e^-
\,.
\end{equation}
By eliminating the $d$'s, we have
\begin{equation}\label{acC0}
  {\cal L}^{IIA}_{gf, 2}  =
 - 2 {\rm i}\, \psi^T_+  \, \frac{C (1 - \Gamma_{11})}{2} \, {\mathcal  M}_-^{-1}\,  \frac{(1 - \Gamma_{11})}{2} \psi_-\,.
 \end{equation}
and after some simple algebra, one gets
\begin{equation}\label{acC}
  {\cal L}^{IIA}_{gf, 2}   =
 - \frac{1}{2 \, e}\, \Phi^T_+  \Big(C_4 \otimes \bar\tau^7 + {\rm i}\, C_4\, \gamma^5 \otimes \bfone_6 \Big) \Phi_-\,.
 \end{equation}
Finally summing the $B^{[2]}$ part and the contribution
of the ghost fields we have the quadratic part of the fermionic action
\begin{eqnarray}\label{compACT}
 {\cal L}^{IIA}_{gf, 2}
 &-& \frac{1}{\, e}\, \Phi^T_+  \Big(\frac{1}{4} C_4 \otimes \bar\tau^7 - \frac{\rm i}{2} \, C_4\, \gamma^5 \otimes \bfone_6 \Big) \Phi_- \, e^+ \wedge e^- \\
&+& \left(\frac{1}{2} w^T_-
\Big(C_4 \otimes \bfone_6 -  \gamma^5 \otimes \bar\tau^7\Big)  \nabla_+ \lambda
- \frac{1}{2}  w^T_+
\Big(C_4 \otimes \bfone_6 +  \gamma^5 \otimes \bar\tau^7\Big)  \nabla_- \lambda
\right) e^+\wedge e^-
\,. \nonumber
\end{eqnarray}
Notice that the matrices $(C_4 \otimes \bfone_6 \pm  \gamma^5 \otimes \bar\tau^7)$ are projectors and
by using the result of the appendix (\ref{inversion}), $\bar\tau^7 _{AB} =
\overline\eta_A \tau^7 \eta_B = K_{AB}$, we see that the projectors couple the 4-d chirality to the eigenspaces of $K_{AB}$.

The third line of eq. (\ref{expA}) vanishes on our background by showing that
$$\mathcal{S}_{L/R}\mathcal{M}=\mathcal{S}_{R}\mathcal{S}_{L}\mathcal{M}=0\,.$$
Using the formulae in \cite{D'Auria:2008ny} one can easily verify
that $\mathcal{S}\mathcal{M}=0$ since the BRST transformation of the
RR field strengths $G_{\underline{ab}},\,G_{\underline{abcd}}$
vanishes as a consequence of the fact that, on our background,
$\chi=\mathcal{D}_{\underline{a}} \chi=\rho_{\underline{ab}}=0$.
The vanishing of $\mathcal{S}_{R}\mathcal{S}_{L}\mathcal{M}=0$, on
the other hand, follows from the properties
$\mathcal{S}\chi=\mathcal{S}\mathcal{D}_{\underline{a}}\chi=\mathcal{S}\rho_{\underline{ab}}=0$,
which must hold for consistency and which can be recast, on our
background, in the following way:
\begin{eqnarray}
\mathcal{S}\chi &=&
\mathcal{N}\,\lambda=0\,\,,\,\,\,\mathcal{S}\mathcal{D}_{\underline{a}}\chi=-\mathcal{N}\,\mathcal{M}\,
\Gamma_{\underline{a}}\,\lambda=0\,\,,\,\,\,
\mathcal{S}\rho_{\underline{ab}}=\left(\mathcal{M}\,\Gamma_{[\underline{a}}\,\mathcal{M}\,\Gamma_{\underline{b}]}-
\frac{1}{4}\,R_{\underline{ab},\underline{cd}}
\,\Gamma^{\underline{cd}}\right)\,\lambda=0\,.\nonumber
\end{eqnarray}
The above equations are satisfied in virtue of the ansatz
(\ref{pureans}) and the Killing spinor equations (\ref{gravino}),
(\ref{ditalino}).

The last line can be computed and we get
\begin{eqnarray}\label{4term}
{\mathcal L}^{\rm IIA}_{gf, 4} &=&  \frac{1}{4} \overline{w}_+
{\mathcal M}_- \Gamma_{\underline{ab}} w_- \overline\lambda_L
\Gamma^{[\underline{a}} {\mathcal M}_+ \Gamma^{\underline{b}]}
\lambda_R\,.
\end{eqnarray}

By simple algebra, (\ref{4term}) can be decomposed in terms of the
eigenspaces of ${\mathcal K}_{AB}$ and of given chiralities so as to
get the expected form of the action
\begin{equation}\label{4termB}
{\mathcal L}^{\rm IIA}_{gf, 4}  = R^{ab,cd} \, N_{ab,+}\,  N_{cd,-}
+ R^{I~~J~~}_{~K~~L} N_{I,+}^{~K} N_{J,-}^{~L}
\end{equation}
where $R^{ab,cd}$ is the $\mathrm{AdS}_4$ Riemann tensor and $
R^{I~~J~~}_{~K~~L} $ is the Riemann tensor for ${\mathbb P}^3$. The
bilinears $ N_{ab},   N_{I,+}^{~K} $ are the Lorentz generators of
$\mathrm{SO}(1,3)$ and of $\mathrm{U}(3)$ of the subgroup of the
coset $\mathrm{Osp}(6|4)/\mathrm{SO}(1,3) \times \mathrm{U}(3)$.
They can be written compactly in $4\oplus6$ notation as follows
\begin{eqnarray}
N_{\underline ab, +} \equiv \bar{w}_+ \Gamma_{{ab}} \lambda _R &=&
-\frac{i}{8}  \left( \overline{w}_{I,+}
   \left({\bf 1}+\gamma _5\right) \gamma_{ab} \lambda^I +
   \overline{w}^I_- \left({\bf 1}-\gamma
   _5\right) \gamma _{{ab}} \lambda_I \right) \nonumber \\
N_{\underline ab, -} \equiv \bar{w}_- \Gamma_{\underline{ab}}
\lambda _L &=& -\frac{i}{8}  \left( \overline{w}^I_-
   \left({\bf 1}+\gamma _5\right) \gamma_{ab} \lambda_I +
   \overline{w}_{I,-} \left({\bf 1}-\gamma
   _5\right) \gamma _{{ab}} \lambda^I \right)
   \end{eqnarray}

Notice that the specific form of the action is dictated by the
invariance under the gauge symmetry of the subgroup
$\mathrm{SO}(1,3) \times \mathrm{U}(3)$ and by the pure spinor
conditions. By using the decomposition as in \cite{Fre:2008qc} it is
easy to perform the Fierz identities. Even if the result is written in a different notation, 
the equivalence with \cite{Bonelli:2008us} can be easily checked.

\section{Conclusion}

We have shown how to derive the pure spinor sigma model for the
background $AdS_4 \times {\mathbb P}^3$. Using the formulation
provided in \cite{D'Auria:2008ny}, we have specified all tensors
appearing in the general action and we have compared with the
formulation derived in   \cite{Fre:2008qc}. The action is the
classical starting point form where to compute higher order
corrections in $\alpha'$. Of course, one can repeat the work done in
the case of $AdS_5 \times S^5$ and check the conformal invariance.
We leave this work to a future work.

%%%%%%%%%%%%%%%%%%%%%%%%%%%%%%%%%%%%%%%%%%%%%%%

\newpage
\appendix
\section{$D=6$ and $D=4$ gamma matrix bases}
\label{spinorbasis} In the discussion of the $\mathrm{AdS_4} \times
\mathbb{P}^3$ compactification we need to consider the decomposition
of the $D=10$ gamma matrix algebra into the tensor product of the
$\so(6)$ clifford algebra times that of $\so(1,3)$. In this section
we discuss and explicit basis for the $\so(6)$ gamma matrix algebra
using that of $\so(7)$. Conventionally we identify the $7$-matrix
$\tau_7$ with the chirality matrix in $d=6$.
\subsection{$D=6$ Clifford algebra}
\label{d7spinorbasis} In this paper, the indices
$\alpha,\beta,\dots$ run on six values and denote the vector indices
of $\so(6)$. In order to discuss the gamma matrix basis we introduce
$\so(7)$ indices
\begin{equation}
  \overline{\alpha} = \alpha, 7
\label{extendedalpha}
\end{equation}
which run on seven values and we define
the Clifford algebra with negative metric:
\begin{equation}
  \left\{ \tau_{\overline{\alpha}} \, , \, \tau_{\overline{\beta}}\right\}
   \, = \, - \delta_{\overline{\alpha \beta}}
\label{taualgebra}
\end{equation}
This algebra is satisfied by the following, real, antisymmetric matrices:
{\scriptsize
\begin{eqnarray*}
  \begin{array}{ccccccc}
    \tau_1 & = & \left(
\begin{array}{llllllll}
 0 & 0 & 0 & 0 & 0 & 0 & 0 & 1 \\
 0 & 0 & 1 & 0 & 0 & 0 & 0 & 0 \\
 0 & -1 & 0 & 0 & 0 & 0 & 0 & 0 \\
 0 & 0 & 0 & 0 & 0 & 0 & 1 & 0 \\
 0 & 0 & 0 & 0 & 0 & -1 & 0 & 0 \\
 0 & 0 & 0 & 0 & 1 & 0 & 0 & 0 \\
 0 & 0 & 0 & -1 & 0 & 0 & 0 & 0 \\
 -1 & 0 & 0 & 0 & 0 & 0 & 0 & 0
\end{array}
\right) &; & \tau_2 & = & \left(
\begin{array}{llllllll}
 0 & 0 & -1 & 0 & 0 & 0 & 0 & 0 \\
 0 & 0 & 0 & 0 & 0 & 0 & 0 & 1 \\
 1 & 0 & 0 & 0 & 0 & 0 & 0 & 0 \\
 0 & 0 & 0 & 0 & 0 & 1 & 0 & 0 \\
 0 & 0 & 0 & 0 & 0 & 0 & 1 & 0 \\
 0 & 0 & 0 & -1 & 0 & 0 & 0 & 0 \\
 0 & 0 & 0 & 0 & -1 & 0 & 0 & 0 \\
 0 & -1 & 0 & 0 & 0 & 0 & 0 & 0
\end{array}
\right) \\
\end{array}
\end{eqnarray*}
\begin{eqnarray*}
  \begin{array}{ccccccc}
    \tau_3 & = & \left(
\begin{array}{llllllll}
 0 & 1 & 0 & 0 & 0 & 0 & 0 & 0 \\
 -1 & 0 & 0 & 0 & 0 & 0 & 0 & 0 \\
 0 & 0 & 0 & 0 & 0 & 0 & 0 & 1 \\
 0 & 0 & 0 & 0 & -1 & 0 & 0 & 0 \\
 0 & 0 & 0 & 1 & 0 & 0 & 0 & 0 \\
 0 & 0 & 0 & 0 & 0 & 0 & 1 & 0 \\
 0 & 0 & 0 & 0 & 0 & -1 & 0 & 0 \\
 0 & 0 & -1 & 0 & 0 & 0 & 0 & 0
\end{array}
\right) & ; & \tau_4 & = & \left(
\begin{array}{llllllll}
 0 & 0 & 0 & 0 & 0 & 0 & -1 & 0 \\
 0 & 0 & 0 & 0 & 0 & -1 & 0 & 0 \\
 0 & 0 & 0 & 0 & 1 & 0 & 0 & 0 \\
 0 & 0 & 0 & 0 & 0 & 0 & 0 & 1 \\
 0 & 0 & -1 & 0 & 0 & 0 & 0 & 0 \\
 0 & 1 & 0 & 0 & 0 & 0 & 0 & 0 \\
 1 & 0 & 0 & 0 & 0 & 0 & 0 & 0 \\
 0 & 0 & 0 & -1 & 0 & 0 & 0 & 0
\end{array}
\right)  \
\end{array}
\end{eqnarray*}
\begin{eqnarray*}
  \begin{array}{ccccccc}
   \tau_5 & = & \left(
\begin{array}{llllllll}
 0 & 0 & 0 & 0 & 0 & 1 & 0 & 0 \\
 0 & 0 & 0 & 0 & 0 & 0 & -1 & 0 \\
 0 & 0 & 0 & -1 & 0 & 0 & 0 & 0 \\
 0 & 0 & 1 & 0 & 0 & 0 & 0 & 0 \\
 0 & 0 & 0 & 0 & 0 & 0 & 0 & 1 \\
 -1 & 0 & 0 & 0 & 0 & 0 & 0 & 0 \\
 0 & 1 & 0 & 0 & 0 & 0 & 0 & 0 \\
 0 & 0 & 0 & 0 & -1 & 0 & 0 & 0
\end{array}
\right) & ; & \tau_6 & = & \left(
\begin{array}{llllllll}
 0 & 0 & 0 & 0 & -1 & 0 & 0 & 0 \\
 0 & 0 & 0 & 1 & 0 & 0 & 0 & 0 \\
 0 & 0 & 0 & 0 & 0 & 0 & -1 & 0 \\
 0 & -1 & 0 & 0 & 0 & 0 & 0 & 0 \\
 1 & 0 & 0 & 0 & 0 & 0 & 0 & 0 \\
 0 & 0 & 0 & 0 & 0 & 0 & 0 & 1 \\
 0 & 0 & 1 & 0 & 0 & 0 & 0 & 0 \\
 0 & 0 & 0 & 0 & 0 & -1 & 0 & 0
\end{array}
\right)  \
\end{array}
\end{eqnarray*}
\begin{equation}
  \begin{array}{ccc}
    \tau_7 & = & \left(
\begin{array}{llllllll}
 0 & 0 & 0 & 1 & 0 & 0 & 0 & 0 \\
 0 & 0 & 0 & 0 & 1 & 0 & 0 & 0 \\
 0 & 0 & 0 & 0 & 0 & 1 & 0 & 0 \\
 -1 & 0 & 0 & 0 & 0 & 0 & 0 & 0 \\
 0 & -1 & 0 & 0 & 0 & 0 & 0 & 0 \\
 0 & 0 & -1 & 0 & 0 & 0 & 0 & 0 \\
 0 & 0 & 0 & 0 & 0 & 0 & 0 & 1 \\
 0 & 0 & 0 & 0 & 0 & 0 & -1 & 0
\end{array}
\right)  \
  \end{array}
\label{tauexplicit}
\end{equation}
}\subsection{$D=4$ $\gamma$-matrix basis and spinor identities}
\label{d4spinorbasis} In this section we construct a basis of
$\so(1,3)$ gamma matrices such that it explicitly realizes the
isomorphism $\so(2,3)  \sim \sym (4,\mathbb{R})$ with the
conventions used in the main text. Naming $\sigma_i$ the standard
Pauli matrices:
\begin{equation}
  \sigma_1 =\left( \begin{array}{cc}
    0 & 1 \\
    1 & 0 \
  \end{array}\right) \quad ; \quad \sigma_2 =\left( \begin{array}{cc}
    0 & -{\rm i} \\
    {\rm i} & 0 \
  \end{array}\right) \quad ; \quad \sigma_3 =\left( \begin{array}{cc}
    1 & 0 \\
    0 & -1 \
  \end{array}\right)
\label{paulini}
\end{equation}
we realize the $\so(1,3)$ Clifford algebra:
\begin{equation}
  \left\{ \gamma_a \, , \, \gamma_b \right\} \, = \, 2 \, \eta_{ab}
  \quad ; \quad \eta_{ab} \, = \, \mbox{diag} \left( + , - , - , - \right)
\label{d4clif}
\end{equation}
by setting:
\begin{equation}
  \begin{array}{ccccccc}
    \gamma_0 & = &  \sigma_2 \, \otimes \, \mathbf{1}  & ; & \gamma_1 & = & {\rm i} \, \sigma_3  \,
    \otimes \, \sigma_1\\
    \gamma_2 & = & {\rm i} \sigma_1 \, \otimes \, \mathbf{1} & ; &
    \gamma_3 & = & {\rm i} \sigma_3 \, \otimes \, \sigma_3 \\
    \gamma_5 & = & \sigma_3 \, \otimes \, \sigma_2 & ; & \mathcal{C} & = & {\rm i} \sigma_2 \,
    \otimes \, \mathbf{1}
  \end{array}
\label{gammareala}
\end{equation}
where $\gamma_5$ is the chirality matrix and $\mathcal{C}$ is the
charge conjugation matrix. Making now reference to eq.s
(\ref{OmandHmat}) and (\ref{ortosymp}) of the main text we see that
the antisymmetric matrix entering the definition of the
orthosymplectic algebra, namely $\mathcal{C}\, \gamma_5$ is the
following one:
\begin{equation}
  \mathcal{C}\, = \, {\rm i} \left(
 \begin{matrix}{ 0 & 0 & 1 & 0 \cr 0 & 0 & 0 & 1 \cr -1 & 0 & 0 & 0 \cr
    0 & -1 & 0 & 0 \cr  }\end{matrix} \right)\,, \hspace{2cm}
  \mathcal{C}\, \gamma_5 \, = \epsilon = \, {\rm i} \left(
 \begin{matrix}{ 0 & 0 & 0 & 1 \cr 0 & 0 & -1 & 0 \cr 0 & 1 & 0 & 0 \cr
    -1 & 0 & 0 & 0 \cr  }\end{matrix} \right)
\label{Chat}
\end{equation}
namely it is proportional, through an overall ${\rm i}$-factor, to a real
completely off-diagonal matrix. On the other hand all the generators
of the $\so(2,3)$ Lie algebra, \textit{i.e.} $\gamma_{ab}$ and
$\gamma_a\, \gamma_5$ are real, symplectic $4 \times 4$ matrices.
Indeed we have
\begin{equation}
  \begin{array}{ccccccc}
    \gamma_{01} & = &\left( \begin{matrix}{ 0 & 0 & 0 & -1 \cr 0 & 0 & -1 & 0 \cr 0 &
    -1 & 0 & 0 \cr -1 & 0 & 0 & 0 \cr  }\end{matrix}\right)  & ; & \gamma_{02} & = & \left(
 \begin{matrix}{ 1 & 0 & 0 & 0 \cr 0 & 1 & 0 & 0 \cr 0 & 0 &
    -1 & 0 \cr 0 & 0 & 0 & -1 \cr  }\end{matrix}\right) \\
    \null & \null & \null & \null & \null & \null & \null \\
    \gamma_{12} & = &\left( \begin{matrix}{ 0 & 0 & -1 & 0 \cr 0 & 0 & 0 & 1 \cr
    -1 & 0 & 0 & 0 \cr 0 & 1 & 0 & 0 \cr  }\end{matrix}\right) & ; & \gamma_{13} & = &\left( \begin{matrix}{ 0 & 0 & 0 & -1 \cr 0 & 0 &
    -1 & 0 \cr 0 & 1 & 0 & 0 \cr 1 & 0 & 0 & 0 \cr  }\end{matrix}\right)\\
    \null & \null & \null & \null & \null & \null & \null \\
    \gamma_{23} & = & \left( \begin{matrix}{ 0 & 1 & 0 & 0 \cr
    -1 & 0 & 0 & 0 \cr 0 & 0 & 0 & 1 \cr 0 & 0 & -1 & 0 \cr  }\end{matrix}\right) & ; & \gamma_{34} & = &\left(
 \begin{matrix}{ 0 & 0 & 1 & 0 \cr 0 & 0 & 0 & -1 \cr
    -1 & 0 & 0 & 0 \cr 0 & 1 & 0 & 0 \cr  }\end{matrix}\right) \\
    \null & \null & \null & \null & \null & \null & \null \\
    \gamma_{0} \,\gamma_5 & = & \left( \begin{matrix}{ 0 & 0 & 0 & 1 \cr 0 & 0 & -1 & 0 \cr 0 & 1 & 0 & 0 \cr
    -1 & 0 & 0 & 0 \cr  }\end{matrix}\right) & ; & \gamma_{1} \,\gamma_5 & = &\left(
 \begin{matrix}{ -1 & 0 & 0 & 0 \cr 0 & 1 & 0 & 0 \cr 0 & 0 &
    -1 & 0 \cr 0 & 0 & 0 & 1 \cr  }\end{matrix}\right) \\
    \null & \null & \null & \null & \null & \null & \null \\
    \gamma_{2} \,\gamma_5 & = & \left( \begin{matrix}{ 0 & 0 & 0 & -1 \cr 0 & 0 & 1 & 0 \cr 0 & 1 & 0 & 0 \cr
    -1 & 0 & 0 & 0 \cr  }\end{matrix}\right)& ; & \gamma_{3} \,\gamma_5 & = &\left( \begin{matrix}{
   0 & 1 & 0 & 0 \cr 1 & 0 & 0 & 0 \cr 0 & 0 & 0 & 1 \cr 0 & 0 &
   1 & 0 \cr  }\end{matrix}\right) \
  \end{array}
\label{realgammi}
\end{equation}
On the other hand we find that $\mathcal{C}\gamma_0 = {\rm i} \,
\mathbf{1}$. Hence the Majorana condition becomes:
\begin{equation}
  {\rm i} \, \psi \, = \, \psi^\star
\label{Majorana}
\end{equation}
so that a Majorana spinor is just a real spinor multiplied by an
overall phase $\exp \left [- i \frac \pi 4\right ]$.
\par
These conventions being fixed let $\chi_x$ ($x=1,\dots,4$) be a set of
(commuting) Majorana spinors normalized in the following way:
\begin{equation}
  \begin{array}{lclcl}
    \chi_x & = & \mathcal{C}\, \overline{\chi}_x^T & ; & \mbox{Majorana condition} \\
    \overline{\chi}_x \, \gamma_5 \, \chi_y & = & {\rm i} \, \left( \mathcal{C}\, \gamma_5\right) _{xy} & ; &
    \mbox{symplectic normal basis} \
  \end{array}
\label{festoso}
\end{equation}
Then by explicit evaluation we can verify the following Fierz
identity:
\begin{equation}
  \ft 12 \, \gamma^{ab} \, \chi_z \, \overline{\chi}_x \, \gamma_5 \,
  \gamma_{ab} \, \chi_y \, - \, \gamma_a \, \gamma_5 \, \chi_z \,
  \overline{\chi}_x \, \gamma_a \, \chi_y \, = \, - \, 2{\rm i} \, \left[
  \left( C\gamma_5\right) _{zx} \, \chi_y \, + \, \left( C\gamma_5\right) _{zy} \,
  \chi_x \right]
\label{firzusd4x1}
\end{equation}
Another identity which we can prove by direct evaluation is the
following one:
\begin{eqnarray}
& \overline{\chi}_x \, \gamma_5 \gamma_{ab} \, \chi_y \, \overline{\chi}_z \, \gamma^b \, \chi_t
\, - \, \overline{\chi}_z \, \gamma_5 \gamma_{ab} \, \chi_t \, \overline{\chi}_x \, \gamma^b \, \chi_y
= &\nonumber\\
&  {\rm i}\left( \overline{\chi}_x \, \gamma_a \, \chi_t \, \left(  \mathcal{C}\,
\gamma_5\right)_{yz} \, + \, \overline{\chi}_y \, \gamma_a \, \chi_t \, \left( \mathcal{C}\,
\gamma_5\right)_{xz}  + \overline{\chi}_x \, \gamma_a \, \chi_z \, \left( \mathcal{C}\,
\gamma_5\right)_{yt} \, + \, \overline{\chi}_y \, \gamma_a \, \chi_z \, \left( \mathcal{C}\,
\gamma_5\right)_{xt}\right) &\nonumber\\
\label{firzusd4x2}
\end{eqnarray}

\par
Finally let us mention some relevant formulae for the derivation of
the compactification. With the above conventions we find:
\begin{equation}
  \gamma_0\,\gamma_1 \, \gamma_2\, \gamma_3 \, =  \, {\rm i} \,
  \gamma_5
\label{g5ide}
\end{equation}
and if we fix the convention:
\begin{equation}
  \epsilon_{0123} \, = \, + \, 1
\label{conveps}
\end{equation}
we obtain:
\begin{equation}
  \ft {1}{24} \, \epsilon^{abcd} \, \gamma_a \, \gamma_b \, \gamma_c
  \, \gamma_d \, = \, - \, {\rm i} \, \gamma_5
\label{epsgamma}
\end{equation}

%%%%%%%%%%%%%%%%%%%%%%%%%%%%%%%%%%%%%%%%%%

\section{An $\so(6)$ inversion formula}
\label{inversion}
In order to discuss the conversion of supergravity forms into MC
forms of the supercoset a key role is played by an inversion formula
which we utilize in the main text and we discuss in this appendix.
Let us define the following set of $6 \times 6$ matrices:
\begin{eqnarray}
\bar{\tau}_{AB}^\alpha & \equiv & \eta_A^T \, \tau^\alpha \, \eta_B \nonumber\\
\bar{\tau}_{AB}^{\alpha\beta} & = & \eta_A^T \, \tau^{\alpha\beta} \,
\eta_B \nonumber\\
{K}_{AB} & = &
\mathcal{K}_{AB}=\frac{1}{2}\,\mathcal{K}_{\alpha\beta}\,\bar{\tau}_{AB}^{\alpha\beta}\,.
\label{definizie}
\end{eqnarray}
where $\eta_A$ are the $6$ killing internal killing spinors and
$\tau$ denote the $1$-index and $2$-index $\so(6)$ gamma-matrices.
By construction the barred $\bar\tau$.s are antisymmetric $6 \times 6$
matrices, hence $\so(6)$ generators in the fundamental representation
just as the {K\" ahler} form $K$. Counting these matrices we find
that they are $6+15+1$, namely $22$, which is too much as a set of
independent generators of $\so(6)$. This means that there must be
linear dependences. By calculating traces of these matrices we find
that the $6$ matrices $\bar{\tau}^\alpha$ are linear independent and
orthogonal to the $15$ ,  $\bar{\tau}^{\alpha\beta}$, and to the unique $K$ while among these
latter $16$ matrices  only $9$ are linear independent.
\par
This observation is important for the following reason. When we write
the following formulae:
\begin{eqnarray}
\Delta \mathcal{B}^{\alpha } & = & - \ft 18 \, {\bar \tau}^{\alpha}_{AB} \, \mathcal{A}^{AB} \nonumber\\
\Delta \mathcal{B}^{\alpha\beta } & = & \ft e4 \, {\bar \tau}^{\alpha\beta}_{AB} \,
\mathcal{A}^{AB}\,  - \, \ft e4 \, {\mathcal{K}}^{\alpha\beta} \, K_{AB} \, \mathcal{A}^{AB}
\label{pinocchio}
\end{eqnarray}
we are actually decomposing the $\so(6)$ connection
$\mathcal{A}^{AB}$ along an over-complete basis of $15 + 6 = 21$ generators of
$\so(6)$, which is obviously a well defined operation.\par
It is interesting to establish the inverse formula, namely to
express the original connection $\mathcal{A}^{AB}$ in terms of the
over complete set of objects $\Delta B^{\alpha }$ and $\Delta B^{\alpha\beta
}$. The inverse formula can be established by means of direct
calculation in the explicit $\tau$-matrix basis we have chosen and we
find what follows:
\begin{equation}
  \mathcal{A}_{AB} \, = \left( - \, 2 \, \Delta \mathcal{B}^{\alpha } \, {\bar
  \tau}_\alpha \, + \, \ft{1} {4 \, e} \, \Delta \mathcal{B}^{\alpha\beta
} \, {\bar \tau}_{\alpha \beta } - \, \ft{1} {4 \, e} \, \Delta B^{\alpha\beta
} \, {\mathcal{K}}_{\alpha \beta } \, K\right)_{AB}
\label{gonzalo}
\end{equation}

%%%%%%%%%%%%%%%%%%%%%%%%%%%%%%%%%%%%%%%%%%%%%%%

\newpage
\newpage

%%%%%%%%%%%%%%%%%%%%%%%%%%%%%%%%%%%%%%%%%%%%%%

\end{document}